\documentclass[amstex,epsf,amssymb,usenatbib]{mn2e}
\usepackage{epsf}
\usepackage{amsmath,amssymb,graphicx}
\title{A stability analysis of radiative shocks in the presence of a transverse magnetic field}

\author[B. Ramachandran \& M.D. Smith]
       {Babulakshmanan Ramachandran\thanks{E-mail: brc@arm.ac.uk}
\& Michael D. Smith\thanks{E-mail: mds@arm.ac.uk} \\
Armagh Observatory, College Hill, Armagh BT61 9DG, Northern Ireland
}

\date{Accepted .....
      Received ..... ;
      in original form .....}

\pagerange{\pageref{firstpage}--\pageref{lastpage}}
\pubyear{2005}

\begin{document}
\maketitle
\label{firstpage}
\begin{abstract} Radiative shock waves may be subject to a global thermal instability in 
which the cooling layer and shock front undergo growing resonant oscillations. For strong 
hydrodynamic shocks, the presence of the overstability depends on the temperature and density 
indices of power-law cooling functions and the specific heat ratio, $\alpha$, $\beta$ and 
$\gamma$, respectively. Here, we investigate the stabilising influence of a transverse magnetic field
by introducing the shock Alfv\'en number, $M_a$ as a fourth parameter. We thus investigate 
the stability criteria for both molecular and atomic shocks under a wide range of conditions. 
In particular, we find that all molecular shocks in which the cooling increases with the temperature
($\alpha > 0$) are stabilised to the first four modes if $M_a < 20$ ($\beta = 2$). For $\alpha = -0.5$,
the first overtone remains stable only for   $M_a < 8$. We conclude that molecular shocks
in the interstellar medium are probably stabilised by a transverse magnetic field unless
exceptional circumstances arise in which the cooling strongly increases as the gas cools. 

\end{abstract}


\section{Introduction}              
\label{intro}

To interpret observations of an astrophysical shock we need to know if it
can be compared to steady state models or if it is subject to an instability.
In fact, the cooling immediately downstream of a shock front may lead to an overstability
of the entire radiative shock wave \citep{1981ApJ...245L..23L}. For hydrodynamic flows, a 
linear analysis yields conditions for growing oscillations as
as well their particular frequencies \citep{1982ApJ...261..543C}, and many analytical and 
numerical studies have been presented, as recently summarised by 
\cite[][hereafter, Paper~1]{2005MNRAS.357..707R}.
However, the growth of the oscillations will 
be damped by a magnetic field. A transverse field not only reduces the immediate post shock 
temperature and compression but could also completely stabilise the cooling layer. The
first question is: how strong does the field have to be to ensure stability?
This question has been answered for the important case of fast shocks into atomic gas
\citep{1993ApJ...413..176T,1997ApJ...487..728K}. Here, we extend these results to 
include  power-law cooling functions pertaining
to other interstellar conditions: a molecular gas and a medium 
in which the energy levels relevant to the cooling are maintained in 
local thermodynamic equilibrium. The conclusions take on added significance 
through the latest campaigns to explore infrared and submillimetre
regimes with spectroscopic methods (e.g. with Spitzer and Herschel).
 
A linear analysis of plane-parallel radiative shocks with a transverse field was performed
by \cite{1993ApJ...413..176T}. They assumed a  specific heat ratio ($\gamma$) of 5/3 and a
cooling function ($\Lambda$) $\propto$  $\rho^{2}~T^{\alpha}$, where $\rho$ is the 
density and $T$ is the temperature. Only the index $\alpha$
determines the stability in the absence of a magnetic field provided the shock 
is strong. The latter assumption eliminates the Mach number, $M$, from the problem. Here,
$M = u_{in}/c_s$ where $u_{in}$ is the shock speed and $c_s$ is the upstream sound speed. 

On the other hand, the Alfv\'en number $M_a = u_{in}/v_a$, where $v_a$ is the
upstream Alfv\'en speed, is a second variable when a significant magnetic field is present. 
\cite[][hereafter TD93]{1993ApJ...413..176T} found, as an example,  that even a quite weak 
field ($M_a < 8$) will suppress the growth of all modes of oscillation for $\alpha > 0$. 
For $\alpha = 0.5$, even weaker fields ($M_{a} < 33$) are enough to stabilise a shock.
Their numerical study agreed with the linear analysis and also revealed that for shock 
speeds $v_{s} < 160$~km~s$^{-1}$, radiative shocks occurring in interstellar gas with 
$n_{H} \leq 0.4$~cm$^{-3}$ may be magnetically stabilised. Further simulations by 
\cite{1997ApJ...487..728K}  however, also demonstrated that the typical interstellar 
field may not be sufficient to stabilise shocks if $\alpha < 0$. The fundamental mode 
can generate very large amplitude oscillations. Moreover, even when the fundamental is 
stabilised, the overtones can still produce higher frequency oscillations of substantial amplitude.

The general linear analysis for hydrodynamic flows was extended in Paper~1 to include molecular
shocks. The dependence on three parameters was considered: $\alpha$, $\beta$ and $\gamma$ where
$\beta$ is the density dependence of the cooling: 
\begin{equation}
\Lambda \propto \rho^{\beta}~T^{\alpha}.
\end{equation}
A strong dependence on $\gamma$ was found, with the regime of overstability significantly
reduced for molecular shocks. In particular the fundamental mode grows only for $\alpha < -0.24$)
in the molecular equivalent ($\gamma = 7/5$ and $\beta = 2$) of the atomic case
(for which $\alpha < 0.38$ is required). However, the overtones are more
significant in molecular shocks with, for example, the first overtone growing for
$\alpha < 0.66$. Building on this work, we introduce the Alfv\'en number, here as 
the fourth parameter, to determine the influence of the magnetic field.

In the molecular shocks discussed here, we presume that the fraction of mass in ions 
is sufficiently high so that ion-magnetosonic waves cannot propagate ahead of
the shock front. Otherwise, if the fraction of ions would be low, the magnetic field will drift 
through the neutral molecules, a process termed ambipolar diffusion. The friction
between the ions and neutrals may then replace the molecular viscosity,
changing the nature of the shock front. We discuss this restriction in \S~\ref{results}. 
Other dynamic instabilities are then
possible in such Continuous Shocks (as well as in Jump Shocks, see Paper~1 for a summary). 
The Jump Shocks studied here thus require a source
of ionisation that does not cause wholesale molecule dissociation. This may occur
near the edges of molecular clouds partially exposed to the ultraviolet from nearby OB associations,
as proposed for the HH\,90/91 shock \citep{1994A&A...289..256S}. The ions can be dust grains,
molecules or atoms, according to the physical conditions. 

In this paper, we extend the formalism of TD93. 
Starting with the  steady state radiative shock structure, we perturb the shock velocity and  
determine the growth or damping rate in addition to the oscillatory period. We study the one dimensional 
case while ignoring thermal conduction as well as radiation transfer. In the hydrodynamic case, a 
stationary wall of infinite density is assumed to lie downstream of the shock front. In the MHD analysis, 
the density in the downstream gas rises towards a finite constant value. In order to make the
problem tractable, we not only ensure that the temperature continues to fall, reaching zero Kelvin,
but that the temperature perturbation also vanishes at the downstream boundary. 
Then, waves escaping downstream are constrained to be pure Alfv\'en waves, leading to a straightforward
boundary condition. In this respect, it should be noted that the published versions of 
equation (A27) of TD93  (where the minus sign should be replaced by a +) and equation (2) of 
\cite{1997ApJ...487..728K} (where the conditionals should be reversed) are incorrect.

In Paper~1, we undertook  a  linear stability analysis similar to \cite{1996ApJ...458..327I} 
while taking a general $\gamma$ so as to encompass molecular shocks. By also generalising the density 
index, we take into account different physical situations. In particular, 
$\beta = 1$ corresponds to  cases where the energy levels of the atoms or molecules which 
provide the dominant  cooling are in local thermodynamic equilibrium. For example,
this can correspond to H$_2$ cooling in warm gas for densities above $\sim 10^4$~cm$^{-3}$
provided hydrogen atoms act as the main collision partner. The specific
heat ratio $\gamma = 5/3$ remains appropriate for dissociative shocks.
The value  $\gamma = 7/5$ applies to a pure H$_2$ gas while, more
realistically for the interstellar medium, $\gamma = 10/7$ accounts
for the addition of ten per cent helium atoms. 

\begin{figure}
\begin{center} 
\epsfxsize=8.7cm 
\epsfbox{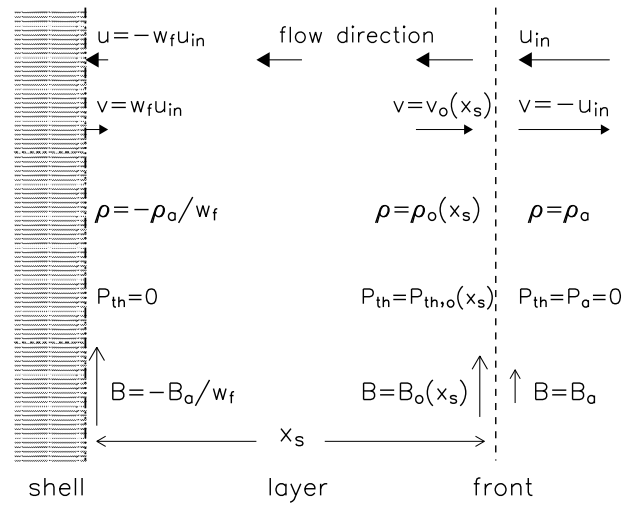}
\caption[]{A sketch of the steady shock configuration. The shock jump
conditions are defined by equations~\ref{rhozero}, \ref{velzero}
and \ref{pzero}; the total compression, $-w_{f}$, is given by equation~\ref{bceq}. 
Note that under strong shock conditions, the upstream thermal pressure is equated to zero.}
\label{steadymag}
\end{center}
\end{figure}
\begin{figure*}
\centering   
	\includegraphics[width=17.0cm]{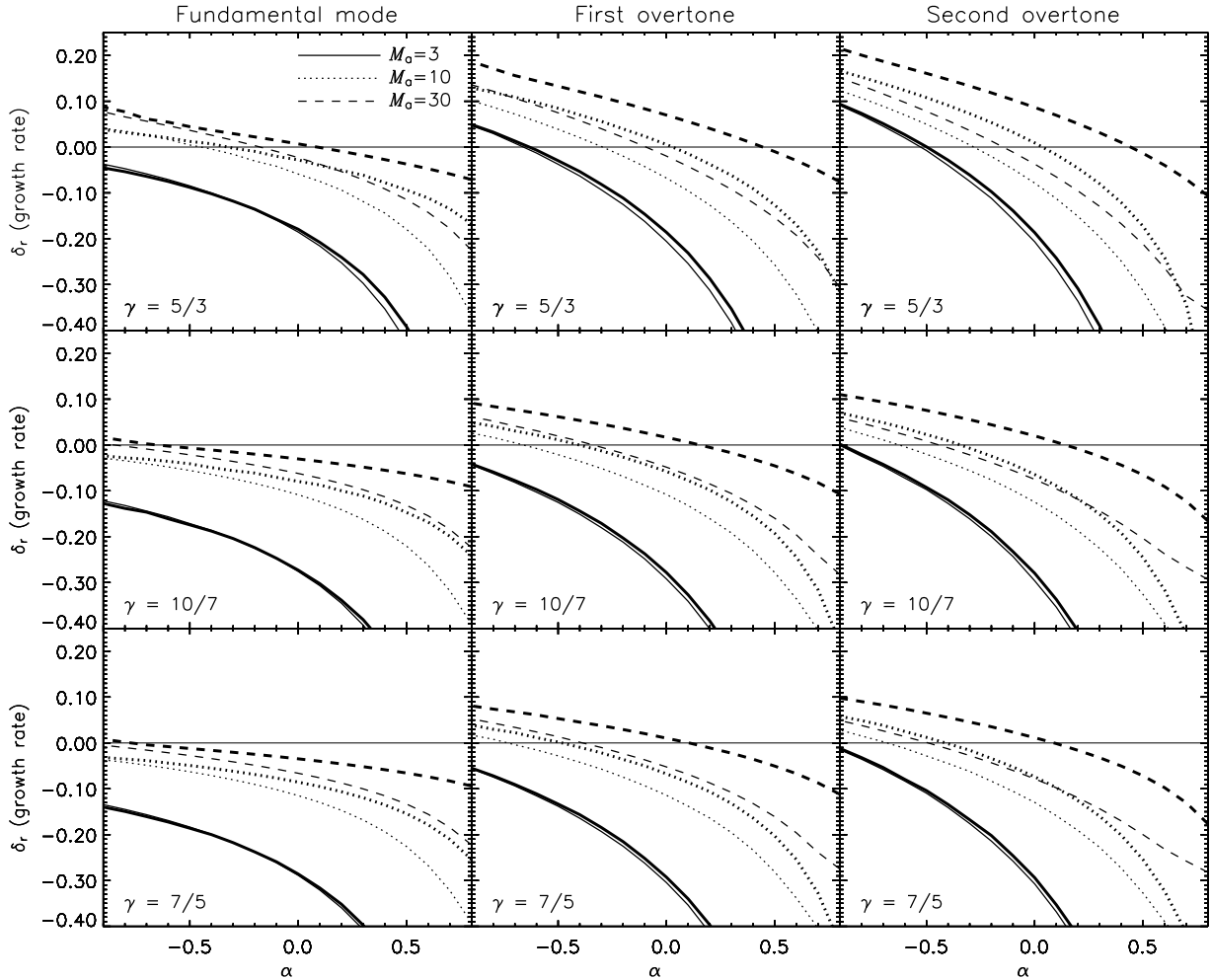} 
\caption{A graphical summary of results for the stability. The growth/damping rates, $\delta_{r}$ plotted against $\alpha$ for the 
fundamental, first and second mode (from left to right), the indicated  different cases of $\gamma$ (from top to bottom), different $\beta$
(thick: $\beta = 2$; thin; $\beta = 1$) and three magnetic field strengths, indicated by the line styles
shown in the top left-hand panel by the shock Alfv\'en numbers, $M_a$.}
\label{growthrates}
\end{figure*} 

\begin{figure*}
\centering   
	\includegraphics[width=17.0cm]{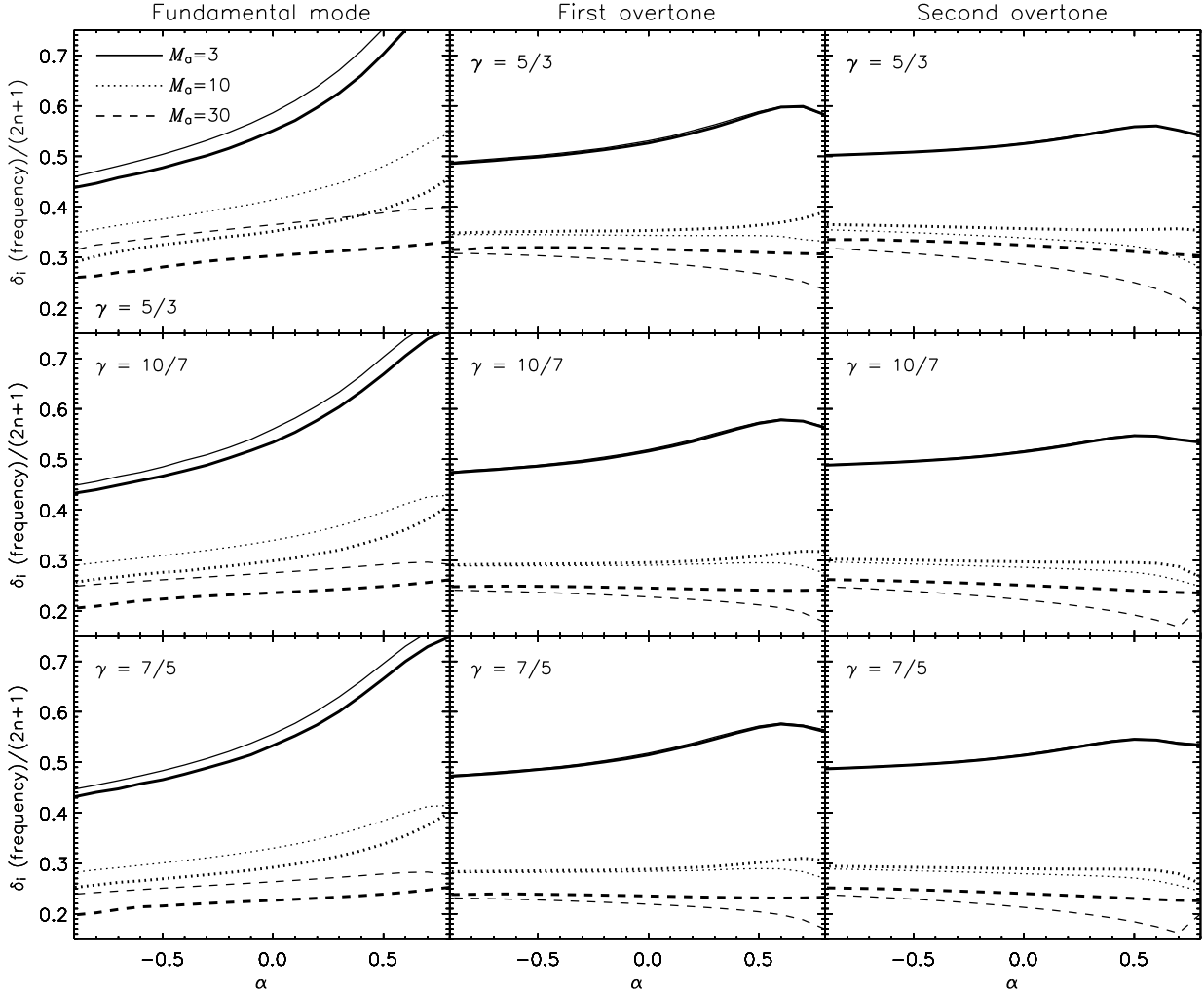} 
\caption{A graphical summary of results for the frequency. The frequency, $\delta_{i}$, plotted here against $\alpha$ has been normalised
by the factor $1/(2n+1)$ where n is the mode number (fundamental being n\,=\,0), which demonstrates that the modes resemble that of a half-open 
organ pipe. Displayed are results for the 
fundamental, first and second mode (from left to right), the indicated  different cases of $\gamma$ (from top to bottom), different $\beta$
(thick: $\beta = 2$; thin; $\beta = 1$) and three magnetic field strengths, indicated by the line styles
shown in the top left-hand panel by the shock Alfv\'en numbers, $M_a$.}
\label{frequencies}
\end{figure*} 

\begin{figure}
\centering                                                                  %
	\includegraphics[width=8.0cm]{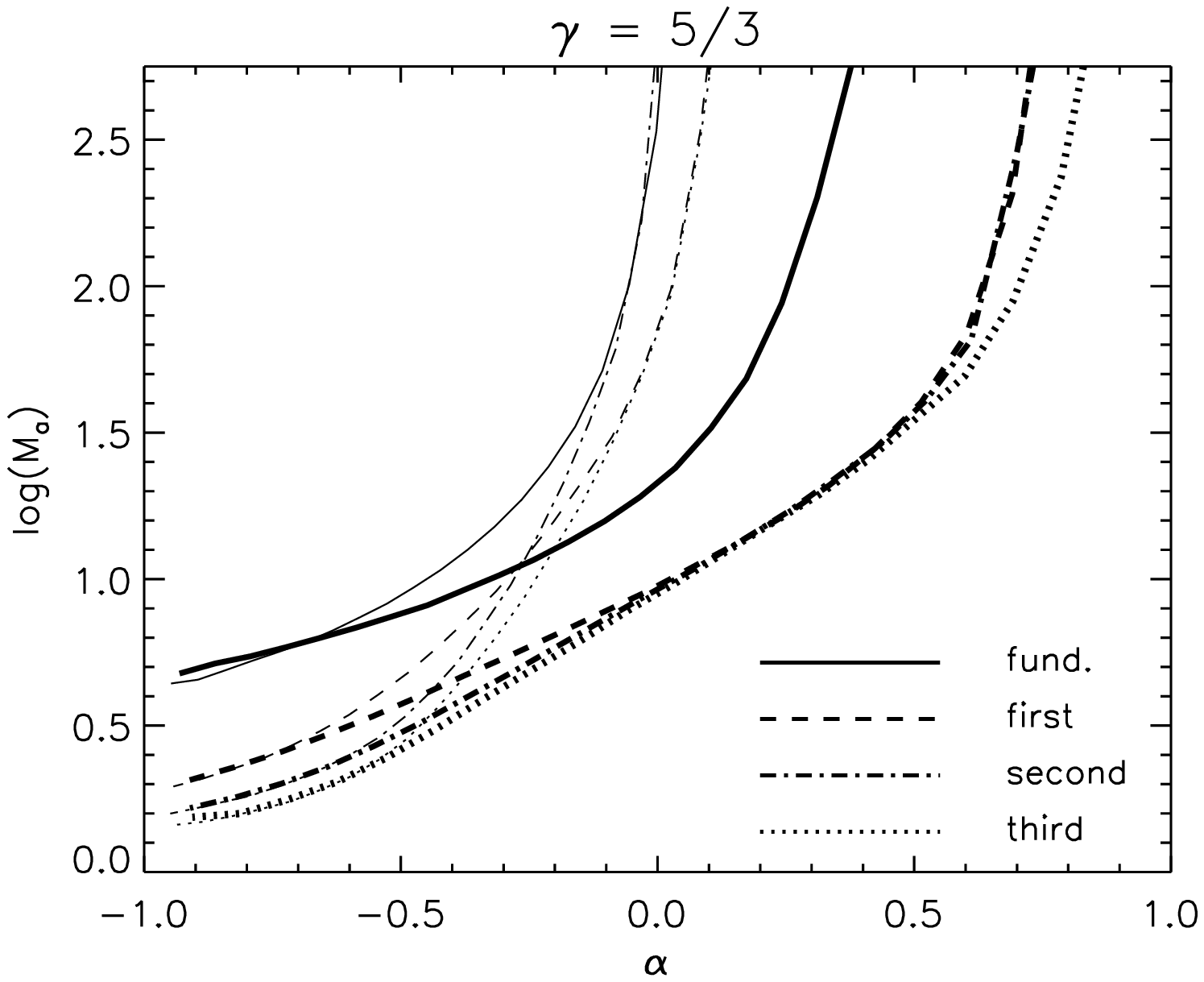} 	
        \includegraphics[width=8.0cm]{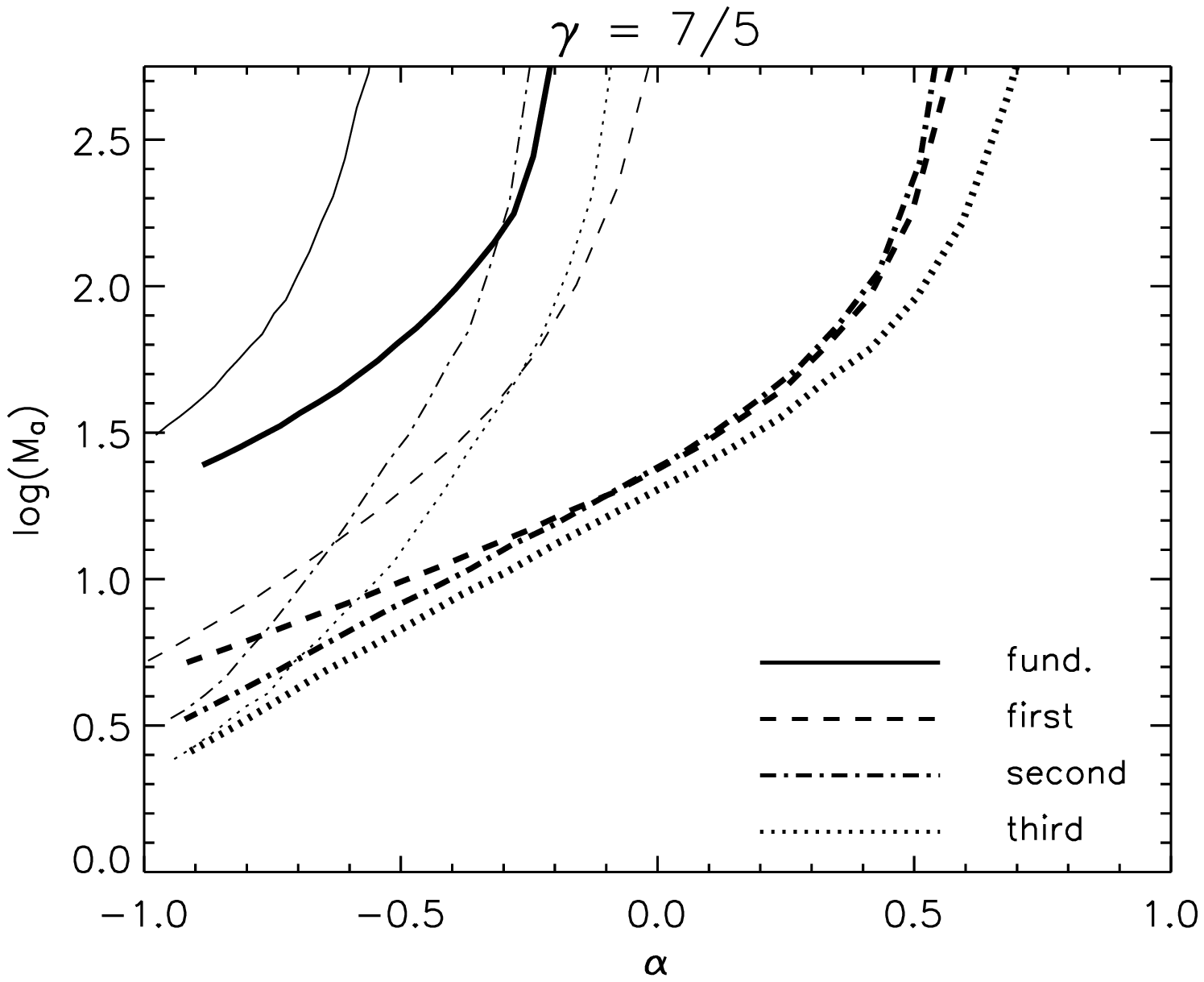}
\caption{The loci of neutral stability for the indicated  values of $\gamma$, modes and
$\beta = 2$ represented by thick lines and $\beta = 1$ by thin lines.
log ($M_{a}$) is plotted against $\alpha$).
The region of instability is above the lines in all cases. 
}
\label{locii}
\end{figure}

\begin{figure}
\centering                                                                  %
	\includegraphics[width=8.0cm]{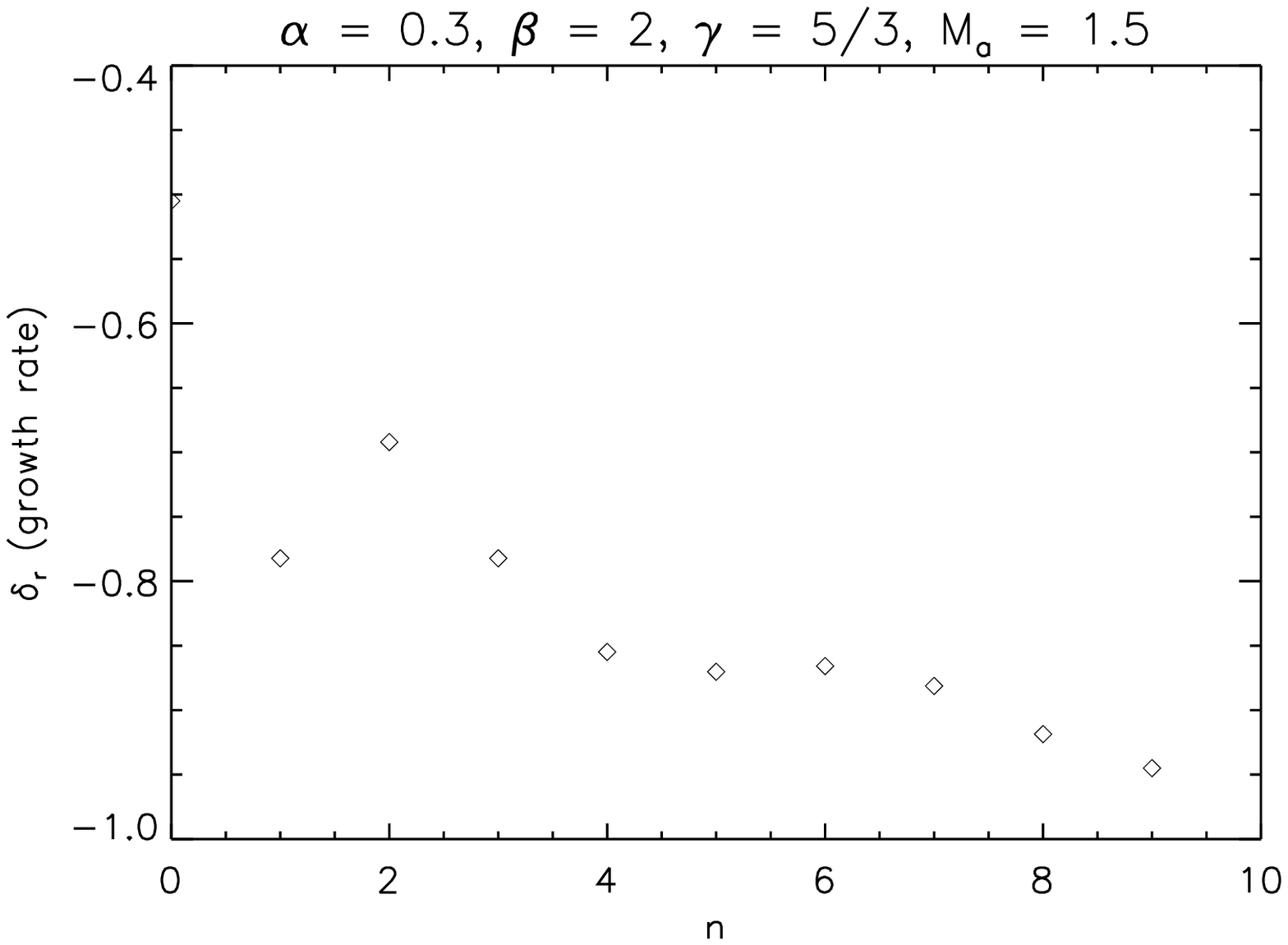} 	
        \includegraphics[width=8.0cm]{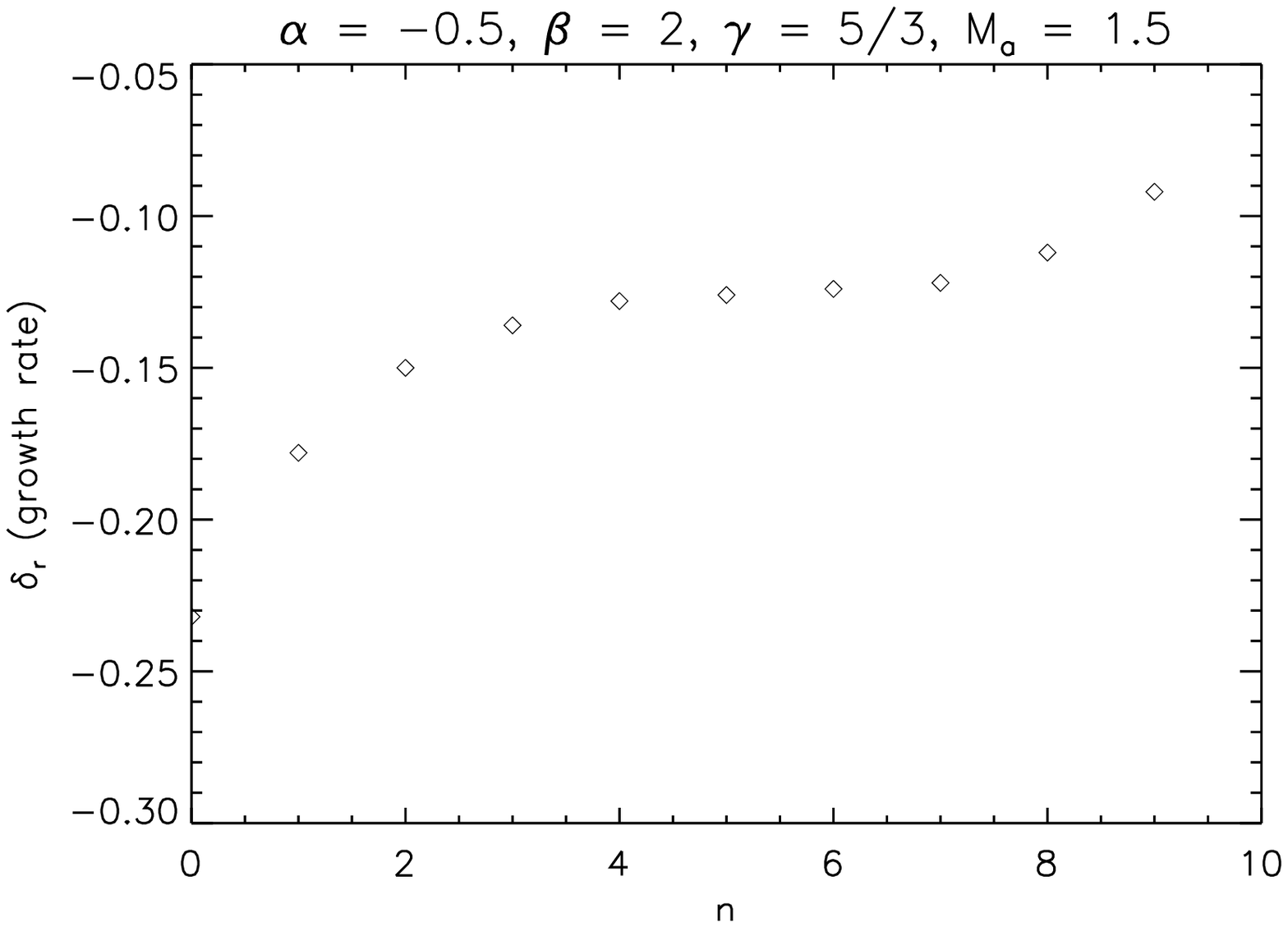}
\caption{The growth rate for various overtones in the presence of a strong transverse magnetic field
for the cases $\alpha = 0.3$ (upper panel) and $\alpha = -0.5$ (lower panel)
on taking $\beta = 2$, $\gamma = 5/3$ and $M_{a}$ = 1.5.}
\label{ndr}
\end{figure}


\section{Formulation of the problem}
\label{method}

\subsection{The equations}

We consider a reference frame in which an ambient gas of uniform  density $\rho_{a}$ and supersonic speed $u_{in}$ passes 
through a stationary shock front at $x = x_s$ from upstream i.e. to the right in  Fig.~\ref{steadymag}, as defined 
by \cite{1982ApJ...261..543C}. The gas cools and collapses before reaching a cold shell at $x = 0$, 
also of uniform density. The approach 
speed $u_{in}$ is defined as a positive quantity. Therefore, the pre-shock gas velocity is $v = -  u_{in}$. In this 
frame, the material in the shell is moving downstream, thus making space for the accumulation of new material. 

Following TD93, we take a perpendicular magnetic field, $B_{a}$, which is frozen into the gas and so increases 
in proportion to the density. The shock Alfv\'en (Mach) number is defined as  $M_a = u_{in}/v_a$ where $v_a$ is the Alfv\'en
speed in the ambient medium. The magnetic pressure in the flow is then
\begin{eqnarray}
\label{eqmagp}
 P_{B} = b \rho^{2},
\end{eqnarray}
where the constant $b$ is
\begin{equation}
   b = \frac {B_{a}^{2}} { 8\pi \rho_{a}^{2} } = v_a^2/2,
\end{equation}
and we define the useful parameter $\theta$ (the parameter $\beta$ of TD93) as 
\begin{equation}
   \theta = \frac {B_{a}^{2}} { 8\pi \rho_{a}u_{in}^{2} } =  1/(2~M_a^2),
\end{equation}
which is the ratio of the magnetic pressure to the upstream ram pressure.

The one dimensional hydrodynamical equations are (e.g. TD93)
\begin{eqnarray}
  \label{eqmass}
   {\partial \rho \over \partial t} +  {\partial ({\rho}v) \over \partial x} &=& 0,  \\
\label{eqmom}
   \rho \left({\partial v \over \partial t} + v{\partial v \over \partial x}\right) 
   + {\partial P \over \partial x} &=& 0
\end{eqnarray}
and
\begin{equation}
\label{eqint}
   {\partial e \over \partial t} +  {\partial (ev) \over \partial x} 
   = - {\partial (Pv) \over \partial x} - \Lambda,  \\
\end{equation}
where $\Lambda$ is the radiated energy loss per unit volume and P is the 
total (magnetic+thermal) pressure related to the internal energy by
\begin{equation}
   e = \frac{P}{\gamma - 1} + \left( 1 -\frac{1}{\gamma - 1}\right)\frac{B^{2}}{8\pi}
     + \frac{1}{2}{\rho}v^2.
\end{equation}
Equations (\ref{eqmass}), (\ref{eqmom}) \& (\ref{eqint}) refer to the conservation 
of mass, momentum and energy, respectively. Eliminating $e$,
\begin{multline}\label{eqenergy}
   {\partial P \over \partial t} + v{\partial P \over \partial x} 
    + \left[\gamma P + b \rho^{2}(2 - \gamma)\right] {\partial v\over \partial x} =\\
   -(\gamma - 1) A \rho^{\beta - \alpha} (P - b\rho^{2})^{\alpha}, 
\end{multline}
where $A$ is a constant and we have used the ideal gas law to eliminate temperature.

\subsection{The steady state}

The steady state solution is denoted by the subscript 0. Equations (\ref{eqmass}) and (\ref{eqmom}) are integrated to yield
\begin{eqnarray}
\label{masscon}
   \rho_{0}v_{0} &=& - \rho_{a}u_{in}\\
\label{momcon}
   P_{0} &=& \rho_{a} u_{in}(v_{0} + u_{in}) + b\rho_{a}^{2}
\end{eqnarray}

\noindent
Equations (\ref{masscon}) \& (\ref{momcon}) are substituted in equation (\ref{eqenergy}) which results in
\begin{eqnarray}
\label{9}
\frac{dv_{0}}{dx} = \frac{C(-\rho_{a}u_{in})^{\beta - 1} \left[- v_{0}^{2}- u_{in}v_{0} +  \rho_{a} b \left( \frac {u_{in}}{v_{0} } - \frac {v_{0}}{u_{in}}\right)   \right]^{\alpha} }{v_{0}^{\beta}\left[v_{0} + \gamma(v_{0} + u_{in}) + \rho_{a} b \left( \frac {\gamma}{u_{in} } + \frac {u_{in}(2 - \gamma)}{v_{0}^{2}}\right)\right]},
\end{eqnarray}
where
\begin{eqnarray}
\label{c}
C = (\gamma - 1)*A. \nonumber
\end{eqnarray}
This equation can be integrated to determine the velocity through the shocked layer.
We now introduce the following variables: 
\begin{eqnarray}
\label{xi}
\xi &=& \frac{x}{x_{s}}\\
\label{w}
w &=& \frac{v_{0}}{u_{in}}.
\end{eqnarray}
Equations (\ref{9}), (\ref{xi}) and (\ref{w}) lead to 
\begin{eqnarray}
\label{10}
\frac{d\xi}{dw} &=& 
\frac { -(-w)^{\beta} \left[w + \gamma(1 + w +\theta) + \frac {(2 - \gamma)\theta} {w^{2}}\right] } { u_{in}^{2\alpha - 3 } C\rho_{a}^{\beta - 1}x_{s}\left(-w - w^{2} - \theta w + \frac{\theta}{w}\right)^{\alpha} }.
\end{eqnarray}

\noindent
The jump condition across the steady shock front can be written as  \citep{1989MNRAS.238..235S}
\begin{eqnarray}
\label{rhozero}
   \rho_{0}(x_{s}) &=& \rho_{a} q,\\
\label{velzero}
   v_{0}(x_{s}) &=& - \frac {u_{in}}{q},\\
\label{pzero}
   P_{0}(x_{s}) &=&  \rho_{a}u_{in}^{2}\left( 1- \frac {1}{q} + \theta \right)
\end{eqnarray}
where the shock front compression is
\begin{eqnarray}
\label{q}
   q&=& \frac{(U + V)}{4\theta (2 - \gamma)},\\
   U&=& -(\gamma - 1 + 2\gamma\theta), \nonumber\\ 
   V&=& \left[ \sqrt {4\gamma\theta(\gamma\theta - (\gamma - 1)) + (\gamma - 1)^{2} + 16\theta}\right]. \nonumber
\end{eqnarray}

\noindent
The boundary condition at $x = 0$ is determined from the fact that the temperature
is zero: the total pressure is equal to the magnetic pressure and the  density remains finite:
\begin{eqnarray}
\label{bcsw}
   P_{0}(x = 0) &=&  P_{B}(x = 0).
\end{eqnarray}
If we denote the velocity of the shell as $v_{0}(x = 0)$,
and also define $w_{f} =  \frac{v_{0}(x  =  0)}{u_{in}}$, we can then write 
from (\ref{eqmagp}) and (\ref{momcon}) the expression 
\begin{eqnarray}
\label{bcswa}
 \rho_{a}u_{in}^{2}(1 + w_{f} + \theta)&=&  \rho_{a}u_{in}^{2}\left(\frac{\theta}{w_{f}^{2}}\right).
\end{eqnarray}

\noindent 
The physical solution to the above cubic equation is 
\begin{eqnarray}
\label{bceq}
 w_{f} &=& \frac {-\theta -\sqrt {(\theta^{2} + 4 \theta)}}{2}.
\end{eqnarray}
Note that $-1/ w_{f}$ is the total compression, as indicated in Fig.~\ref{steadymag}.

\subsection{The set of linear equations}

The shock wave is now perturbed by
\begin{eqnarray}
\label{13}
   \frac{dx_{s}}{dt} &=& v_{s1}e^{\sigma t},
\end{eqnarray}
\noindent
where $\sigma = \sigma_{R} + i\sigma_{I}$ is the frequency and $v_{s1}$ is a real quantity.
\noindent 
The position of the shock may be represented as the real part of
\begin{eqnarray}
\label{14}
x_{s} &=& x_{s0} + x_{s1}e^{\sigma t}
\end{eqnarray}

\noindent
where $x_{s1} = v_{s1}/{\sigma}$. Considering only the terms up to first order:
\begin{eqnarray}
\label{15}
   \xi = \frac{x}{x_{s}} &=& \frac{x}{x_{s0}} \left(1 -\frac{x_{s1}}{x_{s0}}e^{\sigma t} \right)\\
\label{16}
   {\partial \xi \over \partial x} &=&  \frac{1}{x_{s0}}
   \left(1 - \frac{x_{s1}}{x_{s0}}e^{\sigma t}\right)  \\
\label{17}
   {\partial \xi \over \partial t} &=&  -\frac{x x_{s1} \sigma e^{\sigma t}}{x_{s0}^{2}} \\
\label{18}
   \rho &=& \rho_{0}(\xi) + \rho_{1}(\xi)e^{\sigma t}\\
\label{19}
   P &=& P_{0}(\xi) + P_{1}(\xi)e^{\sigma t}\\
\label{20}
   v &=& v_{0}(\xi) + v_{1}(\xi)e^{\sigma t}.
\end{eqnarray}
All the quantities with subscript 1 represent the small perturbed factors.

When the shock is moving, we consider the frame of reference in which the shock is stationary, 
i.e., the upstream velocity is now $-u_{in} - v_{s1}e^{\sigma t}$. The boundary conditions at the 
moving shock wave are obtained by taking the derivatives of equations (\ref{rhozero}), (\ref{velzero}) and 
(\ref{pzero}) with respect to $u_{in}$ to obtain
\begin{eqnarray}
\label{21}
   \rho_{1}(x = x_{s}) &=& 2 v_{s1}\theta q^{2} \frac {\rho_{a}}{u_{in}}\frac{d(1/q)}{d\theta},  \\
\label{22}
   P_{1}(x = x_{s})  &=& 2v_{s1}\rho_{a}u_{in}\left( 1 - \frac{1}{q} + \theta \frac{d(1/q)}{d\theta}\right),\\
\label{23}
   v_{1}(x = x_{s})  &=& v_{s1}\left(2\theta \frac{d(1/q)}{d\theta} - \frac{1}{q} \right) +  v_{s1}.
\end{eqnarray}
where the last term in (\ref{23}) arises because of the transformation of coordinates to the shock frame.

We will also need the derivative of $q$ which is, from equation (\ref{q}),
\begin{eqnarray}
\frac{d(1/q)}{d\theta}  &=& \left[ \gamma + \frac { 4 + \gamma(2\gamma\theta - (\gamma - 1)) } {\sqrt{V} }\right]\nonumber\\ && \times \left [\frac { 2 - \gamma} { 2 - \gamma(\gamma - 1)}\right].
\end{eqnarray}

We then transform to the following variables: 
\begin{eqnarray}
\label{24}
   \zeta &=& \frac{x_{s0}\sigma \rho_{1}}{v_{s1}\rho_{a}},\\
\label{25}
   \pi &=& \frac{P_{1}}{v_{s1}\rho_{a}u_{in}},\\
\label{26}
   \eta &=& \frac{v_{1}}{v_{s1}},\\
\label{27}
   \delta &=& \frac{x_{s0}\sigma}{u_{in}}.
\end{eqnarray}

\noindent
At the shock front, from (\ref{21}) - (\ref{23}), the new variables take the following values:
\begin{eqnarray}
\label{bcxi}
\xi (w =  w_{s})& =& 1,\\
\label{bcrho21}
\zeta(w = w_{s}) &=& 2 \delta\theta q^{2}\frac{d(1/q)}{d\theta},  \\
\label{bcpr22}
\pi(w = w_{s})  &=& 2\left( 1 - \frac{1}{q} + \theta \frac{d(1/q)}{d\theta}\right),\\
\label{bcvel23}
\eta(w = w_{s})  &=& \left(2\theta \frac{d(1/q)}{d\theta} - \frac{1}{q} + 1\right) 
\end{eqnarray}
Substituting (\ref{18}), (\ref{19}) and (\ref{20}) into (\ref{eqmass}),  (\ref{eqmom}) and (\ref{eqenergy}), 
the fluid equations can be rewritten as\\
\begin{eqnarray} 
\label{28}
-\frac {\xi}{w^{2}} + \zeta \frac{d\xi}{dw} + \frac{w}{\delta}\frac{d\zeta}{dw} + 
\frac{\eta}{w^{2}} -\frac{1}{w} \frac{d\eta}{dw} + \frac{\zeta}{\delta} &=& 0,\\
\label{29}
-\xi + \delta \frac{d\xi}{dw}\eta + w\frac{d\eta}{dw} + \eta - \frac{w^{2}}{\delta}\zeta -
 w\frac{d\pi}{dw} &=& 0,\\
\label{30}
D + E &=& F
\end{eqnarray} 
\vskip -0.4cm
\noindent
where
\vskip -0.3cm
\begin{eqnarray}
D &=& \left[-\xi  + \delta \frac{d\xi}{dw}\pi + w\frac{d\pi}{dw} +\eta - \frac{w}{\delta} + 
\pi \gamma \right]\nonumber\\
E &=& \left(-\frac {\xi}{w^{2}} + 
\zeta \frac{d\xi}{dw} + \frac{w}{\delta}\frac{d\zeta}{dw} + \frac{\eta}{w^{2}} 
-\frac{1}{\delta w}  + \frac{\zeta}{\delta} \right)\nonumber\\ &&\times \left(\gamma (w + 1 + \theta) +\frac {(2 - \gamma)\theta} {w^{2}} \right)w - \frac{2(2 - \gamma)\zeta \theta}{w \delta} \nonumber\\
F &=&\left[w + \gamma (1+w +\theta) +\frac {(2 - \gamma)\theta}{w^{2}}\right]\nonumber\\ &&\times \left[\alpha \pi  - (\beta - \alpha)(1 + w + \theta)\frac {\zeta w}{\delta} + (\beta + \alpha) \frac{\zeta \theta}{\delta w}\right] \nonumber\\&&\times \left[w + 1 + \theta - \frac{\theta}{w^2}\right]^{-1}. \nonumber
\end{eqnarray} 
\noindent
The quantities $\zeta$, $\pi$ and $\eta$ are complex eigenfunctions. We employ the subscript $r$ to denote the real 
component and $i$ stands for the imaginary part for each of the above quantities. The quantity $\delta$ is 
a complex number with the sign of the  real part, $\delta_{r}$, indicating  
the instability (+ve value) or stability (-ve value) of a mode.
The quantity $\delta_{i}$ is interpreted as the eigenfrequency (in units of ($u_{in}/x_{s0}$)). 
The equations for the real and complex quantities are provided in Appendix~\ref{differential}.

\subsection{The integration process}

While integrating the above equations we find that there are numerical problems when $\alpha \leq 0.2$ 
as the temperature approaches zero. To solve this, we follow a prescription suggested by TD93 which is 
to introduce a break in the power law cooling at a temperature $T_c$, at a velocity $w_{c}$, leaving a small 
region near the shell where the cooling function will have $\alpha$ = 0.5. Therefore, the cooling function is 
split into two regimes such that 
\begin{eqnarray}
\label{tc}
\Lambda & =& A\rho^{\beta- \alpha}\left(P - b\rho^{2}\right)^{\alpha} \hskip 1.21cm {\rm at} \hskip 0.2cm w < w_{c}\\  
\Lambda &=& A_{c}\rho^{\beta - \alpha_{c}}\left(P - b\rho^{2}\right)^{\alpha_{c}} \hskip .81cm {\rm at} \hskip 0.2cm w > w_{c}
\end{eqnarray}
Here, $A_{c} = A (P_{c} - b\rho_{c})^{\alpha - \alpha_{c}}$ and we choose $A_{c} = 0.001$ to be consistent with the results of TD93.
To obtain the shock length, we integrate equation~(\ref{10}) 
from $\xi = 1$ to $\xi = 0$ and take into account the two component 
cooling function.

The boundary condition at the shell for a moving shock under the influence of a 
transverse magnetic field is derived in Appendix~\ref{boundary}. It is more complicated 
than in Paper I where the gas settles on the wall.
In the present scenario, the velocity of the gas does not vanish since
the magnetic pressure limits the compression. The boundary 
condition is 
\begin{equation}
  |\pi \sqrt {-\frac{w^{3}}{2\beta}} + \eta| = 0, 
\end{equation}
similar to TD93 but correcting a typographical error in their equation (A27).

For the various combinations of the four free parameters, $\alpha$, $\beta$, $\gamma$ and $\theta$, 
the quantities $\delta_{r}$ and $\delta_{i}$ are determined by imposing the boundary condition at the shell
boundary. We solved the differential equations employing a fourth order Runge-Kutta 
technique for trial values of $\delta_{r}$ and $\delta_{i}$ selected from a grid of points uniformly
covering the complex plane. The combinations that come closest to satisfying the boundary condition for each 
mode determine a new set of grid points with a higher resolution. 

\section{General Results} 
\label{results}


We have calculated the growth rates ($\delta_{r}$) and eigenfrequencies ($\delta_{i}$) for the various  
cases of $\alpha$, $\beta$ and $\gamma$ with weak ($M_{a} = 30$), intermediate ($M_{a} = 10$) and strong  
($M_{a} = 3$) magnetic fields for the fundamental mode as well as the first three overtones. 
These results  are presented in Tables~\ref{g53b1r}\,--\,\ref{g75b2i}. Equivalent tables for the case $\gamma = 10/7$
have been constructed but are not presented here.
We plot the growth rates against $\alpha$ in Fig.~\ref{growthrates} for various modes. 
The major results are as follows.

$\bullet$ We have accurately reproduced the results of TD93 for the specific case $\beta = 2$ and $\gamma = 5/3$ 
(Tables~\ref{g53b2r} \& Table~\ref{g53b2i}) as a specific case in our parametric space.

$\bullet$ The fact that the instability regime becomes increasingly restricted with an increase in the transverse
magnetic field holds for all values of $\alpha$, $\beta$ and $\gamma$.

$\bullet$ The fundamental remains the most restricted mode to be unstable. The field has a similar
stabilising effect on all the modes. 

$\bullet$ The lower value for the density index $\beta$ provides more stability. However, for high magnetic fields
(solid lines in  Fig.~\ref{growthrates}), the stability criterion becomes independent of
$\beta$. This is interpreted as the cushion effect of the field which limits the compression. 

$\bullet$ With decreasing $\gamma$, the instability regime is more restricted for all magnetic field strengths.
Whether the molecular gas is diatomic ($\gamma = 7/5$) or contains helium  ($\gamma = 10/7$)
does not significantly alter the instability regime.

$\bullet$ The frequency of the fundamental decreases strongly as $\alpha$ decreases for all magnetic field strengths.
The frequency of the overtones generally tend to a constant value as $\alpha$ decreases. Note that in
Fig.~\ref{frequencies} the frequencies of the modes have been normalised by a factor of
$2n+1$ where n is the mode number for display purposes and so demonstrates that, for all magnetic field
strengths, the resonant frequencies are analogous to the acoustical modes in a half-open organ pipe. 

The loci of neutral stability in the $\log (M_a)$ -- $\alpha$ space, which separate
stable and unstable regimes (for fixed $\beta$ and $\gamma$), are displayed in Fig.~\ref{locii}.
Whereas an Alfv\'en number of 8 will suppress the growth of all modes of oscillation for $\alpha > 0$ in the atomic
shocks ($\gamma = 5/3, \beta = 2$), we require just $M_a < 20$ for the equivalent molecular case. Hence
molecular shocks tend to be considerably more stable. On the other hand, the  Alfv\'en number of 8 would be 
sufficient to stabilise even the first overtone for molecular shocks with $\alpha = -0.5$.

Given the trends in Fig.~\ref{locii}, it is not clear that
higher order modes possess wider instability ranges. 
In fact, from inspection of the tables (in the appendix),
the growth rate tends to be lower for higher
overtones for $\alpha > 0.0$ especially as the magnetic field strength increases, whereas for $\alpha < 0$, however,
there is an increasing trend in the growth rate. Here,
we have investigated these trends for low $M_a$ shocks
to determine if the magnetic field will stabilise
shocks entirely.
As shown in Fig.~\ref{ndr}, the growth rate continues to decrease
as the mode number is increased for $\alpha = 0.3$ and $M_a = 1.5$,
taking an atomic shock for illustration. In contrast, the growth rate
displays the opposite trend for the case with negative $\alpha$. In other words,
the magnetic field may not provide complete stability for  cooling with
negative $\alpha$. However, the significance of high order modes
are probably highly restricted and a multi-dimensional analysis is necessary
to determine any observational consequences.

The present analysis applies to both atomic and molecular shocks and
assumes that the magnetic field is parallel to the shock front. However,
in a sufficiently strong shock the molecules will dissociate immediately following 
the shock jump. In fact, dissociative cooling may dominate the cooling function.
Given a molecular medium with Alfv\'en speed of 1~km~s$^{-1}$  \citep{1999ApJ...520..706C}
and a speed limit for molecular shocks of 40~km~s$^{-1}$ at low densities (under
$\sim 10^{4}$~cm$^{-3}$) and $\sim 24$~km~s$^{-1}$ at high cloud densities \citep{1994MNRAS.266..238S},
the maximum Alfv\'en number for molecular shocks with $\gamma = 7/5$ is then in the range 24 -- 40. 

A maximum value for the magnetic field should also exist, above which  ambipolar diffusion dominates 
the shock physics. The limit is more difficult to calculate since it depends
on the ionisation fraction, density and cooling function. In fact, the limitation is
much stronger on the ion fraction rather than the magnetic field.
In molecular clouds which are well shielded from the extreme ultraviolet, the ion fraction, $\chi$, is 
extremely low and ambipolar diffusion
dominates. On the other hand, the ion fraction may reach high values near massive stars,
stellar ouflows or within bow shocks with high-speed apices, supplying ionising radiation.
As estimated by \cite{1990MNRAS.242..495S}, ion fractions exceeding $\sim 10^{-5} B_{-3}/n_{6}^{3/2}$ (H$_2$O cooling)
or  $\sim 2~10^{-6} B_{-3}/n_{6}^{1/2}$ (H$_2$ cooling) are required to ensure a J-type
shock where $B_{-3} = B/(10^{-3}$~G) and $n_{6} = n/(10^6 {\rm cm}^{-3})$. Thus,
the Alfv\'en speed must not exceed $\sim 5~10^5n_6\chi$ (H$_2$O cooling)~km~s$^{-1}$ (H$_2$O cooling)
or $\sim 10^5\chi$~km~s$^{-1}$ (H$_2$ cooling).

\section{Conclusions: Molecular Shock}
\label{conclusions}

We have found the stable and unstable regions for radiative shocks in the
$\alpha$, $\beta$ and $\gamma$ parameter space for various magnetic field strengths
and in the field-$\alpha$ parameter space for specific $\beta$ and $\gamma$ values.

For the fundamental mode, molecular shocks are considerably more stable than atomic shocks.
For the overtones, molecular shocks are only moderately more stable.
The magnetic field, however, has a strong stabilising influence on all these modes, decreasing all the growth
rates by similar factors (see Fig.~\ref{growthrates}).

The field strength enters the stability condition through the shock Alfv\'en number. All
cooling functions in which the cooling increases with temperature are stable for $M_a < 20$.
However, for  $M_a > 22$, the first overtone can be unstable, potentially leading to 
oscillations with non-linear amplitudes. Hence, we need to estimate the Alfv\'en number
for typical shocks in various environments. 

Unfortunately, in molecular clouds, the Alfv\'en speed is a very uncertain parameter. 
It is even difficult to get an objective view on the 
magnetic field strength alone \citep{1999ApJ...526..279P} 
although derived Alfv\'en speeds generally lie within the range 0.5 - 5~km~s$^{-1}$ \citep{1999ApJ...520..706C}.
Jump shocks followed by cooling zones can arise without
molecular dissociation for shock speeds as high as 40~km~s$^{-1}$ at low densities
but $\sim 24$~km~s$^{-1}$ at high cloud densities \citep{1994MNRAS.266..238S}. 
As discussed in \S~\ref{results}, a minimum ion fraction is also required
to maintain a frozen-in magnetic field. Hence, a quite fast shock in a diffuse
molecular cloud is the most favourable state for the appearance of oscillations.

The temperature dependence of molecular cooling functions generally correspond to
values of $\alpha$ well above zero and so indicates stable shocks. However,
besides temperature and density, shock cooling may depend on the chemistry. The rate of
formation of trace molecules within the cooling layer can increase the cooling rate as the
temperature falls. In particular, the propagation of warming but non-ionising shocks into 
cool atomic gas may show trace molecule formation and increasing cooling within a gas with  
$\gamma = 5/3$. However, because of the low shock speeds, these shocks would be stabilised 
if the magnetic field were transverse. Hence, it can be that oblique non-ionising shock waves 
into atomic gas represent the rather restrictive conditions for shock instability within the
cool components of the interstellar medium. 

Although  it is well known how an oblique field alters the nature of a steady shock, it is not
clear how it alters the stability conditions. It should also be noted that the magnetic field may
help stabilise shocks even when the field is parallel to the shock provided the Alfv\'en number is less
than the Mach number, a condition expected to be satisfied in molecular clouds 
\citep{1993A&A...272..571S}. Such an analysis, as well as
multi-dimensional numerical simulations of radiative shocks, are still to be performed.

\section*{Acknowledgments}

Research at Armagh Observatory is funded by
the Department of Culture, Arts and Leisure, Northern
Ireland. BR is extremely grateful to Sathya Sai Baba for 
inspiration and also thanks Professor S. Kandaswami
for encouragement and advice. 

\bibliography{magshock}

\appendix
\section{The differential equations}
\label{differential}

>From Eqs. (\ref {25}), (\ref {26}) and (\ref {27}), we get six coupled 
first order equations which are
\begin{eqnarray} 
\label{31}
\frac{d\eta_{r}}{dw} &=& \frac { \alpha \pi_{r} \delta^{2} +(\delta_{r}\zeta_{r} + \delta_{i}\zeta_{i})\left( \frac { (\beta + \alpha)\theta}{w} - w(\beta - \alpha)(1 + w + \theta)\right)}{(1 + w + \theta - \frac{\theta}{w^{2}})\delta^{2}}  \nonumber\\ && +  \frac{2\xi \delta^{2} - 2\eta_{r}  \delta^{2} - \gamma\pi_{r} \delta^{2} +(w^{2} +\frac{2(2 - \gamma)\theta}{w}) (\zeta_{r}\delta_{r} + \zeta_{i}\delta_{i})}{(w + \gamma (w + 1 + \theta) + \frac{(2 - \gamma)\theta}{w ^{2}})\delta^{2}}\nonumber\\ && -\frac{d\xi}{dw}\left[\frac{\delta_{r}(\eta_{r} + \pi_{r}) - \delta_{i}(\eta_{i} + \pi_{i})}{(w + \gamma (w + 1 + \theta) + \frac{(2 - \gamma)\theta}{w ^{2}})}\right]+ \frac{\delta_{r}}{\delta^{2}}\\
\label{32}
\frac {d\eta_{i}}{dw} &=&  \frac { \alpha \pi_{i} \delta^{2} +(\delta_{r}\zeta_{i} - \delta_{i}\zeta_{r})\left( \frac { (\beta + \alpha)\theta}{w} - w(\beta - \alpha)(1 + w + \theta)\right)}{(1 + w + \theta - \frac{\theta}{w^{2}})\delta^{2}}  \nonumber\\ && +  \frac{- 2\eta_{i}  \delta^{2} - \gamma\pi_{i} \delta^{2} +(w^{2} +\frac{2(2 - \gamma)\theta}{w}) (\zeta_{i}\delta_{r} - \zeta_{r}\delta_{i})}{(w + \gamma (w + 1 + \theta) + \frac{(2 - \gamma)\theta}{w ^{2}})\delta^{2}}\nonumber\\ && -\frac{d\xi}{dw}\left[\frac{\delta_{i}(\eta_{r} + \pi_{r}) + \delta_{r}(\eta_{i} + \pi_{i})}{(w + \gamma (w + 1 + \theta) + \frac{(2 - \gamma)\theta}{w ^{2}})}\right]- \frac{\delta_{i}}{\delta^{2}}
\end{eqnarray}
\begin{eqnarray}
\label{33}
\frac {d\pi_{r}}{dw} &=& \frac {-\xi}{w} + \frac {(\delta_{r}\eta_{r} - \delta_{i}\eta_{i})}{w} \frac {d\xi}{dw} + \frac {\eta_{r}}{w} + \frac {d\eta_{r}}{dw}\nonumber\\ && - \frac{w(\zeta_{r}\delta_{r} + \zeta_{i}\delta_{i})}{\delta^{2}}
\end{eqnarray}
\begin{eqnarray}
\label{34}
\frac {d\pi_{i}}{dw} &=& \frac {(\delta_{i}\eta_{r} + \delta_{r}\eta_{i})}{w} \frac {d\xi}{dw} + \frac {\eta_{i}}{w} + \frac {d\eta_{i}}{dw}\nonumber\\ && - \frac{w(\zeta_{i}\delta_{r} - \zeta_{r}\delta_{i})}{\delta^{2}}
\end{eqnarray}
\begin{eqnarray}
\label{35}
\frac {d\zeta_{r}}{dw} &=& \frac {\xi\delta_{r}}{w^{3}} - \frac {d\xi}{dw}\frac{\zeta_{r}\delta_{r}}{w}  + \frac {\delta_{r}}{w^{2}} \frac {d\eta_{r}}{dw} - \frac {\eta_{r}\delta_{r}}{w^{3}}\nonumber\\ && - \frac {\delta_{r}(\zeta_{r}\delta_{r} + \zeta_{i}\delta_{i})}{w\delta^{2}} +\frac {d\xi}{dw}\frac{\zeta_{i}\delta_{i}}{w} - \frac {\delta_{i}}{w^{2}} \frac {d\eta_{i}}{dw}\nonumber\\ && + \frac {\eta_{i}\delta_{i}}{w^{3}} + \frac {\delta_{i}(\zeta_{i}\delta_{r} - \zeta_{r}\delta_{i})}{w\delta^{2}}\\
\label{36}
\frac {d\zeta_{i}}{dw} &=& \frac {\xi\delta_{i}}{w^{3}} - \frac {d\xi}{dw}\frac{\zeta_{i}\delta_{r}}{w}  + \frac {\delta_{r}}{w^{2}} \frac {d\eta_{i}}{dw} - \frac {\eta_{i}\delta_{r}}{w^{3}}\nonumber\\ && - \frac {\delta_{r}(\zeta_{i}\delta_{r} - \zeta_{r}\delta_{i})}{w\delta^{2}} - \frac {d\xi}{dw}\frac{\zeta_{r}\delta_{i}}{w} + \frac {\delta_{i}}{w^{2}} \frac {d\eta_{r}}{dw}\nonumber\\ && - \frac {\eta_{r}\delta_{i}}{w^{3}} - \frac {\delta_{i}(\zeta_{r}\delta_{r} + \zeta_{j}\delta_{i})}{w\delta^{2}}.
\end{eqnarray}

\section{Boundary condition at the shell}
\label{boundary}

Following TD93, we exploit the condition that  the temperature 
tends to zero at the shell. This implies that the only pressure at the shell is 
due to the magnetic field. To extract the boundary condition, we represent the physical 
quantities as  
\begin{eqnarray}
\rho &=& \rho_{0}(x) + \rho_{1}(x)e^{i(ft + kx)}\\
\label{eqp}
P &=& P_{0}(x) + P_{1}(x)e^{i(ft + kx)}\\
v &=& v_{0}(x) + v_{1}(x)e^{i(ft + kx)}
\end{eqnarray}
where $k$ is the wavenumber, which represents the direction of the outgoing wave
towards the shell and $f$ is the frequency. The above equations  are substituted in (\ref{eqmass}) 
and (\ref{eqmom}) to yield
\begin{eqnarray}
\frac{f}{k} = - \left(\frac{\rho_{0} v_{1} v_{0} + P_{1}}{\rho_{0}v_{1}}\right)\\
\frac{f}{k} = - \left(\frac{\rho_{0} v_{1} + \rho_{1}v_{0}}{\rho_{1}}\right)
\end{eqnarray}
Equating the above expressions  leads to 
\begin{eqnarray}
\label{p11}
P_{1} = \frac{\rho_{0}^{2} v_{1}^{2}}{\rho_{1}}
\end{eqnarray}
The magnetic pressure is given by 
\begin{eqnarray}
\label{eqmpp}
P = \frac{B^{2}}{8\pi}
\end{eqnarray}
We now  express the magnetic field in terms of density for example, \citep{1989MNRAS.238..235S} as 
\begin{eqnarray}
    \frac {B}{B_{a}} = \frac {\rho}{\rho_{a}}
\end{eqnarray}
where $B$ and $\rho$ are the post-shock magnetic field and density. Substituting 
(\ref{eqp}) in (\ref{eqmpp})
and defining the Alfv\'{e}n velocity as $v_{A} = \sqrt\frac {B^{2}}{4\pi\rho}$,
we get 
\begin{eqnarray}
\label{p12}
P_{1} = v_{A}^{2}\rho_{1}
\end{eqnarray}
Equations (\ref{p11}) and (\ref{p12})  yield the boundary condition at the shell as,
\begin{eqnarray}
\label{eqp1}
   P_{1} = - v_{A}v_{1}\rho_{0}
\end{eqnarray}
The minus sign indicates the direction of the gas flow. Note that the perturbed quantities are 
function of $x$, which itself is a function of $\xi$ from (\ref{15}). The expression  
(\ref{eqp1}) can be written in terms of the non-dimensional quantities as
\begin{eqnarray}
\label{eqdp}
\pi = -\eta  \sqrt\frac {2\theta}{-w_{f}^3}
\end{eqnarray}
 The above quantities are evaluated at the shell boundary or equivalently at $w = w_{f}$.

\section{Tables} 

\begin{table*}
 \caption{Growth rates ($\delta_{r}$) for several  modes and  magnetic field strengths, as measured by the 
Alfv\'en number $M_a$, for $\gamma = 5/3$, $\beta = 1$. }
 \label{g53b1r} 
$$
\begin{array}{lccccccccc} 
\hline
\mathsf{ \,\,\,\,\,}
\hskip 3.5cm \rm fundamental \hskip 4.5cm \rm first \hskip 0.1cm  overtone  \\
\hline
\noalign{\smallskip} \noalign{\smallskip}
	 \mathsf{\alpha \,\,\,\,\,\,\,}
	 \hskip 1.1cm \mathsf{M_{a} = 3}
	 \hskip 1.0cm \mathsf{M_{a} = 10} 
	 \hskip 1.0cm \mathsf{M_{a} = 30}
	 \hskip 1.36cm \mathsf{M_{a} = 3}
 \hskip 1.0cm \mathsf{M_{a} = 10}
 \hskip 1.0cm \mathsf{M_{a} = 30}
     \\
 \hline  
 \noalign{\smallskip}
 \mathsf{-1} \hskip 0.95cm -0.02942 \hskip 1.05cm 0.04903  \hskip 1.06cm 0.08965  \hskip 1.39cm  0.06387  \hskip 1.cm 0.11420  \hskip 1.18cm 0.14986\\ 
 \mathsf{-0.5} \hskip 0.73cm -0.08446  \hskip 1.03cm 0.00567   \hskip 1.07cm 0.03646  \hskip 1.05cm -0.03742   \hskip 0.97cm 0.03890  \hskip 1.18cm 0.07482 \\ 
 \mathsf{0.0} \hskip 0.95cm -0.18538 \hskip 0.69cm -0.05842 \hskip 0.71cm  -0.02296  \hskip 1.02cm -0.20521  \hskip 0.62cm  -0.06762   \hskip 0.82cm  -0.01974 \\ 
 \mathsf{0.3} \hskip 0.95cm   -0.29720  \hskip 0.69cm -0.11769  \hskip 0.71cm -0.07122   \hskip 1.02cm -0.38342  \hskip 0.62cm -0.16282   \hskip 0.82cm -0.09245  \\ 
  \mathsf{0.5} \hskip 0.95cm -0.42465  \hskip 0.69cm  -0.17862  \hskip 0.71cm -0.11462    \hskip 1.02cm -0.57743   \hskip 0.62cm -0.25705     \hskip 0.82cm -0.15414     \\ 
\noalign{\smallskip}
\hline  \noalign{\smallskip} \noalign{\smallskip} \hline
 \mathsf{ \,\,\,\,\,}
 \hskip 3.5cm \rm second  \hskip 0.1cm  overtone \hskip 4.1cm \rm third \hskip 0.1cm  overtone  \\
\hline \noalign{\smallskip} \noalign{\smallskip}
	 \mathsf{\alpha \,\,\,\,\,\,}
	 \hskip 1.1cm \mathsf{M_{a} = 3}
	 \hskip 1.0cm \mathsf{M_{a} = 10} 
	 \hskip 1.0cm \mathsf{M_{a} = 30}
	 \hskip 1.36cm \mathsf{M_{a} = 3}
 \hskip 1.0cm \mathsf{M_{a} = 10}
 \hskip 1.0cm \mathsf{M_{a} = 30}
     \\
 \hline  
 \noalign{\smallskip}
 \mathsf{-1} \hskip 1.25cm  0.11050 \hskip 1.05cm 0.13937   \hskip 1.06cm 0.16695  \hskip 1.39cm  0.13817   \hskip 1.cm 0.16075   \hskip 1.18cm  0.19142 \\ 
 \mathsf{-0.5}\hskip 0.66cm -0.00862 \hskip 1.035cm 0.04990   \hskip 1.07cm  0.07818    \hskip 1.39cm  0.00809   \hskip  1.cm  0.06541 \hskip 1.18cm 0.09936      \\ 
\mathsf{0.0} \hskip 0.90cm -0.20728 \hskip 0.67cm -0.07856 \hskip 0.69cm -0.03713 \hskip 1.02cm -0.21087 \hskip 0.64cm  -0.07366    \hskip 0.84cm -0.02433  \\ 
\mathsf{0.3} \hskip 0.9cm -0.42137  \hskip 0.67cm -0.19620  \hskip 0.69cm -0.12957   \hskip 1.02cm -0.45058  \hskip 0.64cm  -0.20419  \hskip 0.84cm -0.12893  \\ 
\mathsf{0.5} \hskip 0.88cm -0.65377  \hskip 0.69cm -0.31342   \hskip 0.71cm -0.21165    \hskip 1.02cm  -0.70913    \hskip 0.64cm -0.33717   \hskip 0.82cm -0.22760   \\ 
\noalign{\smallskip}
\hline
\noalign{\smallskip}
\end{array}
 $$
\end{table*}
\begin{table*}
 \caption{Growth rates ($\delta_{r}$)  for several  modes and  magnetic field strengths for $\gamma = 5/3$, 
$\beta = 2$. This Table can be compared to Table~1 of TD93.}
 \label{g53b2r} 
 $$
 \begin{array}{lccccccccc}
\hline
 \mathsf{ \,\,\,\,\,}
\hskip 3.5cm \rm fundamental \hskip 4.5cm \rm first \hskip 0.1cm  overtone  \\
\hline \noalign{\smallskip} \noalign{\smallskip}
	 \mathsf{\alpha \,\,\,\,\,\,}
	 \hskip 1.1cm \mathsf{M_{a} = 3}
	 \hskip 1.0cm \mathsf{M_{a} = 10} 
	 \hskip 1.0cm \mathsf{M_{a} = 30}
	 \hskip 1.36cm \mathsf{M_{a} = 3}
 \hskip 1.0cm \mathsf{M_{a} = 10} \hskip 1.0cm \mathsf{M_{a} = 30} \\
 \hline   \noalign{\smallskip}
 \mathsf{-1} \hskip 0.9cm -0.03549  \hskip 1.05cm 0.04610 \hskip 1.06cm 0.09351  \hskip 1.39cm  0.06696  \hskip 1.cm 0.13981  \hskip 1.18cm 0.19781  \\ 
 \mathsf{-0.5}\hskip 0.66cm -0.08877  \hskip 1.05cm  0.01342   \hskip 1.07cm 0.04383   \hskip 1.02cm  -0.02921   \hskip  1.cm 0.08413  \hskip 1.18cm 0.13291     \\ 
 \mathsf{0.0} \hskip 0.90cm  -0.17983  \hskip 0.67cm  -0.02853  \hskip 1.07cm 0.00702  \hskip 1.02cm -0.18560  \hskip 1.03cm  0.00496  \hskip 1.18cm  0.06983   \\ 
 \mathsf{0.3} \hskip 0.9cm -0.27889 \hskip 0.67cm  -0.06263   \hskip 0.72cm   -0.01806  \hskip 1.02cm -0.35272   \hskip 0.63cm -0.06229  \hskip 1.18cm 0.02479  \\ 
 \mathsf{0.5} \hskip 0.9cm -0.39054 \hskip 0.69cm  -0.09283   \hskip 0.71cm  -0.03697    \hskip 1.0cm  -0.53889   \hskip 0.64cm -0.12623 \hskip 0.82cm  -0.00998 \\ 
\noalign{\smallskip}
\hline \noalign{\smallskip} \noalign{\smallskip}
\hline
 \mathsf{ \,\,\,\,\,}
\hskip 3.5cm \rm second  \hskip 0.1cm  overtone  \hskip 4.1cm \rm third \hskip 0.1cm  overtone  \\
\hline \noalign{\smallskip} \noalign{\smallskip}
	 \mathsf{\alpha \,\,\,\,\,\,}
	 \hskip 1.1cm \mathsf{M_{a} = 3}
	 \hskip 1.0cm \mathsf{M_{a} = 10} 
	 \hskip 1.0cm \mathsf{M_{a} = 30}
	 \hskip 1.36cm \mathsf{M_{a} = 3}
 \hskip 1.0cm \mathsf{M_{a} = 10} \hskip 1.0cm \mathsf{M_{a} = 30} \\
 \hline  \noalign{\smallskip}
 \mathsf{-1} \hskip 1.25cm 0.11374 \hskip 1.05cm  0.17466   \hskip 1.06cm  0.22738  \hskip 1.39cm 0.14253    \hskip 1.cm 0.19826  \hskip 1.18cm  0.25109  \\ 
 \mathsf{-0.5}\hskip 1.02cm  0.00067 \hskip 1.04cm 0.10719   \hskip 1.07cm  0.16104   \hskip 1.39cm   0.01897    \hskip  1.cm 0.12425  \hskip 1.18cm 0.18399   \\ 
 \mathsf{0.0} \hskip 0.90cm  -0.18654 \hskip 1.03cm 0.00890 \hskip 1.05cm 0.08662 \hskip 1.02cm -0.18846 \hskip 1.02cm  0.01377   \hskip 1.19cm 0.10306\\ 
 \mathsf{0.3} \hskip 0.9cm -0.38992   \hskip 0.67cm -0.08046  \hskip 1.07cm  0.03131  \hskip 1.02cm -0.41714  \hskip 0.64cm   -0.09122  \hskip 1.18cm  0.04081 \\ 
  \mathsf{0.5} \hskip 0.88cm  -0.61605  \hskip 0.69cm   -0.17289  \hskip 0.71cm -0.01352  \hskip 1.02cm -0.67020 \hskip 0.64cm -0.20545  \hskip 0.82cm  -0.01179  \\ 
\noalign{\smallskip}
\hline
\noalign{\smallskip}
\end{array}
 $$
\end{table*}
\begin{table*}
 \caption{Growth rates ($\delta_{r}$) for several  modes and  magnetic field strengths for $\gamma = 7/5$, $\beta = 1$. }
 \label{g75b1r} 
 $$
 \begin{array}{lccccccccc}
\hline
 \mathsf{ \,\,\,\,\,}
\hskip 3.5cm \rm fundamental \hskip 4.5cm \rm first \hskip 0.1cm  overtone  \\
\hline \noalign{\smallskip} \noalign{\smallskip}
	 \mathsf{\alpha \,\,\,\,\,\,}
	 \hskip 1.1cm \mathsf{M_{a} = 3}
	 \hskip 1.0cm \mathsf{M_{a} = 10} 
	 \hskip 1.0cm \mathsf{M_{a} = 30}
	 \hskip 1.36cm \mathsf{M_{a} = 3}
 \hskip 1.0cm \mathsf{M_{a} = 10} \hskip 1.0cm \mathsf{M_{a} = 30} \\
 \hline   \noalign{\smallskip}
 \mathsf{-1} \hskip 0.9cm  -0.12574 \hskip 0.67cm -0.03009 \hskip 1.06cm   0.00029   \hskip 1.03cm -0.04055  \hskip 1.cm  0.02876 \hskip 1.12cm 0.06090 \\ 
\mathsf{-0.5}\hskip 0.66cm -0.18312  \hskip 0.69cm  -0.06286   \hskip 0.7cm -0.02743  \hskip 1.01cm  -0.13786  \hskip  0.64cm -0.02838  \hskip 1.1cm  0.01337     \\ 
 \mathsf{0.0} \hskip 0.90cm -0.28865 \hskip 0.67cm -0.11479  \hskip 0.71cm -0.06609  \hskip 1.02cm -0.30446  \hskip 0.63cm -0.11369  \hskip 0.75cm  -0.05206    \\ 
 \mathsf{0.3} \hskip 0.9cm -0.40691 \hskip 0.67cm  -0.16801   \hskip 0.71cm  -0.10022    \hskip 1.01cm -0.48460   \hskip 0.64cm  -0.19620  \hskip 0.76cm  -0.10778  \\ 
 \mathsf{0.5} \hskip 0.9cm  -0.53663 \hskip 0.67cm -0.22589     \hskip 0.69cm -0.13324     \hskip 1.03cm  -0.67452    \hskip 0.63cm  -0.28096 \hskip 0.76cm -0.16065  \\ 
\noalign{\smallskip} \hline \noalign{\smallskip} \hline
 \mathsf{ \,\,\,\,\,}
\hskip 3.5cm \rm second  \hskip 0.1cm  overtone \hskip 4.1cm \rm third \hskip 0.1cm  overtone  \\
\hline \noalign{\smallskip} \noalign{\smallskip}
	 \mathsf{\alpha \,\,\,\,\,\,}
	 \hskip 1.1cm \mathsf{M_{a} = 3}
	 \hskip 1.0cm \mathsf{M_{a} = 10} 
	 \hskip 1.0cm \mathsf{M_{a} = 30}
	 \hskip 1.36cm \mathsf{M_{a} = 3}
 \hskip 1.0cm \mathsf{M_{a} = 10} \hskip 1.0cm \mathsf{M_{a} = 30} \\
 \hline  
 \noalign{\smallskip}
 \mathsf{-1} \hskip 1.25cm 0.00362  \hskip 1.0cm  0.04071   \hskip 1.06cm 0.05939  \hskip 1.39cm  0.02880 \hskip 1.cm   0.06460 \hskip 1.2cm 0.08899  \\ 
 \mathsf{-0.5}\hskip 0.64cm  -0.11048 \hskip 0.65cm  -0.02764   \hskip 1.05cm  0.00114   \hskip 1.02cm  -0.09556   \hskip  0.64cm  -0.09275 \hskip 1.22cm  0.02439     \\ 
 \mathsf{0.0} \hskip 0.88cm -0.30800  \hskip 0.65cm -0.13002  \hskip 0.69cm  -0.07824 \hskip 1.02cm -0.31358  \hskip 0.64cm  -0.12335  \hskip 0.84cm -0.06976   \\ 
 \mathsf{0.3} \hskip 0.88cm -0.52493 \hskip 0.65cm   -0.22824  \hskip 0.69cm  -0.14361    \hskip 1.02cm -0.55789  \hskip 0.64cm -0.23731    \hskip 0.84cm -0.15350  \\ 
  \mathsf{0.5} \hskip 0.88cm  -0.74988  \hskip 0.64cm -0.32688    \hskip 0.69cm -0.20005    \hskip 1.02cm -0.80390    \hskip 0.64cm -0.35412  \hskip 0.85cm -0.22737   \\ 
\noalign{\smallskip}
\hline
\noalign{\smallskip}
\end{array}
 $$
\end{table*}
\begin{table*}
 \caption{Growth rates ($\delta_{r}$)  for several  modes and magnetic field strengths for $\gamma = 7/5$,  $\beta = 2$. }
 \label{g75b2r} 
 $$
 \begin{array}{lccccccccc}
\hline
 \mathsf{ \,\,\,\,\,}
\hskip 3.5cm \rm fundamental \hskip 4.5cm \rm first \hskip 0.1cm  overtone  \\
\hline \noalign{\smallskip} \noalign{\smallskip}
	 \mathsf{\alpha \,\,\,\,\,\,}
	 \hskip 1.1cm \mathsf{M_{a} = 3}
	 \hskip 1.0cm \mathsf{M_{a} = 10} 
	 \hskip 1.0cm \mathsf{M_{a} = 30}
	 \hskip 1.36cm \mathsf{M_{a} = 3}
 \hskip 1.0cm \mathsf{M_{a} = 10} \hskip 1.0cm \mathsf{M_{a} = 30} \\
 \hline   \noalign{\smallskip}
 \mathsf{-1} \hskip 0.9cm -0.12850 \hskip 0.67cm -0.02761    \hskip 1.06cm  0.01627   \hskip 1.09cm  -0.03781   \hskip 1.02cm 0.04672  \hskip 1.19cm   0.08856  \\ 
 \mathsf{-0.5}\hskip 0.68cm -0.18572  \hskip 0.67cm -0.04914   \hskip 0.68cm -0.01205     \hskip 1.09cm  -0.13274     \hskip  1.03cm 0.00162 \hskip 1.18cm  0.05346  \\ 
 \mathsf{0.0} \hskip 0.90cm  -0.28557  \hskip 0.68cm  -0.08540 \hskip 0.7cm  -0.03462  \hskip 1.09cm  -0.29240 \hskip 0.64cm  -0.06666  \hskip 1.17cm  0.01054\\ 
 \mathsf{0.3} \hskip 0.9cm -0.39682   \hskip 0.68cm -0.11950  \hskip 0.69cm -0.05177    \hskip 1.09cm -0.46628 \hskip 0.66cm -0.13231   \hskip 0.8cm  -0.02255  \\ 
  \mathsf{0.5} \hskip 0.90cm  -0.51900 \hskip 0.68cm -0.15418    \hskip 0.69cm  -0.06567  \hskip 1.09cm -0.65304  \hskip 0.66cm -0.20231  \hskip 0.8cm -0.05032  \\ 
\noalign{\smallskip}
\hline  \noalign{\smallskip} \noalign{\smallskip}\hline
 \mathsf{ \,\,\,\,\,}
\hskip 3.5cm \rm second  \hskip 0.1cm  overtone \hskip 4.1cm \rm third \hskip 0.1cm  overtone  \\
\hline \noalign{\smallskip} \noalign{\smallskip}
	 \mathsf{\alpha \,\,\,\,\,\,}
	 \hskip 1.1cm \mathsf{M_{a} = 3}
	 \hskip 1.0cm \mathsf{M_{a} = 10} 
	 \hskip 1.0cm \mathsf{M_{a} = 30}
	 \hskip 1.36cm \mathsf{M_{a} = 3}
 \hskip 1.0cm \mathsf{M_{a} = 10} \hskip 1.0cm \mathsf{M_{a} = 30} \\
 \hline  
 \noalign{\smallskip}
 \mathsf{-1} \hskip 1.25cm  0.00628 \hskip 1.05cm  0.06857   \hskip 1.07cm 0.10574    \hskip 1.38cm   0.03269   \hskip 1.cm  0.09028   \hskip 1.18cm 0.13057   \\ 
 \mathsf{-0.5}\hskip 0.65cm -0.10479  \hskip 1.07cm  0.01285  \hskip 1.05cm 0.06522      \hskip 1.02cm   -0.08846    \hskip  1.02cm 0.02882  \hskip 1.18cm  0.08690   \\ 
\mathsf{0.0} \hskip 0.88cm -0.29458  \hskip 0.7cm  -0.07317 \hskip 1.05cm  0.01163  \hskip 1.02cm   -0.29851 \hskip 0.65cm -0.06802    \hskip 1.19cm   0.02653\\ 
\mathsf{0.3} \hskip 0.88cm -0.50449   \hskip 0.69cm  -0.15932  \hskip 0.7cm -0.03224    \hskip 1.02cm  -0.53521  \hskip 0.65cm -0.16798   \hskip 0.8cm -0.02552 \\ 
\mathsf{0.5} \hskip 0.88cm -0.72683  \hskip 0.69cm -0.25388   \hskip 0.71cm -0.07193  \hskip 1.02cm -0.77992   \hskip 0.64cm  -0.27967 \hskip 0.82cm -0.07602    \\ 
\noalign{\smallskip}
\hline
\noalign{\smallskip}
\end{array}
 $$
\end{table*}

\begin{table*}
 \caption{Eigenfrequencies ($\delta_{i}$) for several  modes and magnetic field strengths  for  $\gamma = 5/3$ and $\beta = 1$. }
 \label{g53b1i} 
 $$
 \begin{array}{lccccccccc}
  \noalign{\smallskip}
 \noalign{\smallskip}
\hline
 \mathsf{ \,\,\,\,\,}
 \hskip 3.5cm \rm fundamental 
\hskip 4.5cm \rm first \hskip 0.1cm  overtone 
 \\
\hline
 \noalign{\smallskip}
 \noalign{\smallskip}
	 \mathsf{\alpha \,\,\,\,\,\,\,}
	 \hskip 1.1cm \mathsf{M_{a} = 3}
	 \hskip 1.0cm \mathsf{M_{a} = 10} 
	 \hskip 1.0cm \mathsf{M_{a} = 30}
	 \hskip 1.36cm \mathsf{M_{a} = 3}
 \hskip 1.0cm \mathsf{M_{a} = 10}
 \hskip 1.0cm \mathsf{M_{a} = 30}
     \\
 \hline  
 \noalign{\smallskip}
 \mathsf{-1} \hskip 1.25cm  0.44847 \hskip 1.05cm 0.33935    \hskip 1.06cm 0.30432  \hskip 1.43cm  1.45520 \hskip 1.cm 1.03328   \hskip 1.18cm  0.92059 \\ 
 \mathsf{-0.5} \hskip 1.02cm 0.50430 \hskip 1.055cm    0.37630  \hskip 1.07cm  0.34126 \hskip 1.43cm 1.50546   \hskip 0.984cm 1.03500   \hskip 1.175cm 0.91030  \\ 
 \mathsf{0.0} \hskip 1.25cm  0.58627 \hskip 1.05cm 0.41387  \hskip 1.07cm 0.36464   \hskip 1.42cm 1.59283  \hskip 1.cm  1.03047   \hskip 1.18cm 0.87269  \\ 
 \mathsf{0.3} \hskip 1.25cm  0.67073   \hskip 1.05cm 0.44723  \hskip 1.07cm 0.37841   \hskip 1.42cm 1.68773   \hskip 1.cm  1.02950  \hskip 1.18cm 0.83627  \\ 
  \mathsf{0.5} \hskip 1.25cm 0.75300  \hskip 1.05cm   0.48020  \hskip 1.07cm 0.38821   \hskip 1.42cm 1.76664   \hskip 1.02cm  1.02936     \hskip 1.14cm  0.80303  \\ 
\noalign{\smallskip}
\hline  \noalign{\smallskip}\noalign{\smallskip} \hline
 \mathsf{ \,\,\,\,\,}
 \hskip 3.5cm \rm second \hskip 0.1cm  overtone  
\hskip 4.1cm \rm third \hskip 0.1cm  overtone 
 \\
\hline
 \noalign{\smallskip}
 \noalign{\smallskip}
	 \mathsf{\alpha \,\,\,\,\,\,\,}
	 \hskip 1.1cm \mathsf{M_{a} = 3}
	 \hskip 1.0cm \mathsf{M_{a} = 10} 
	 \hskip 1.0cm \mathsf{M_{a} = 30}
	 \hskip 1.36cm \mathsf{M_{a} = 3}
 \hskip 1.0cm \mathsf{M_{a} = 10}
 \hskip 1.0cm \mathsf{M_{a} = 30}
     \\
 \hline  
 \noalign{\smallskip}
 \mathsf{-1} \hskip 1.25cm 2.50463  \hskip 1.1cm 1.77852  \hskip 1.06cm 1.59824  \hskip 1.43cm 3.55657  \hskip 1.cm  2.51123    \hskip 1.18cm    2.25140 \\ 
 \mathsf{-0.5} \hskip 1.033cm 2.54768  \hskip 1.1cm  1.74493   \hskip 1.07cm  1.53713  \hskip 1.43cm  3.59392    \hskip 0.984cm 2.44672     \hskip 1.18cm   2.14906  \\ 
 \mathsf{0.0} \hskip 1.25cm 2.63290  \hskip 1.1cm  1.69280 \hskip 1.07cm 1.43223   \hskip 1.42cm   3.67699  \hskip 1.cm   2.35191   \hskip 1.18cm  1.98756   \\ 
 \mathsf{0.3} \hskip 1.25cm 2.72930   \hskip 1.1cm   1.65092  \hskip 1.07cm  1.33791   \hskip 1.42cm 3.77286    \hskip 1.cm  2.27393    \hskip 1.18cm   1.84544  \\ 
  \mathsf{0.5} \hskip 1.25cm   2.79538 \hskip 1.1cm   1.60990    \hskip 1.07cm  1.24986   \hskip 1.42cm  3.82144   \hskip 0.98cm    2.19758    \hskip 1.19cm 1.71171    \\ 
\noalign{\smallskip}
\hline
\noalign{\smallskip}
\end{array}
 $$
\end{table*}

\begin{table*}
 \caption{Eigenfrequencies ($\delta_{i}$) for several modes and magnetic field strengths for $\gamma = 5/3$ and $\beta = 2$. }
 \label{g53b2i} 
 $$
 \begin{array}{lccccccccc}
  \noalign{\smallskip}
 \noalign{\smallskip}
\hline
 \mathsf{ \,\,\,\,\,}
 \hskip 3.5cm \rm fundamental 
\hskip 4.5cm \rm first \hskip 0.1cm  overtone 
 \\
\hline
 \noalign{\smallskip}
 \noalign{\smallskip}
	 \mathsf{\alpha \,\,\,\,\,\,\,}
	 \hskip 1.1cm \mathsf{M_{a} = 3}
	 \hskip 1.0cm \mathsf{M_{a} = 10} 
	 \hskip 1.0cm \mathsf{M_{a} = 30}
	 \hskip 1.36cm \mathsf{M_{a} = 3}
 \hskip 1.0cm \mathsf{M_{a} = 10}
 \hskip 1.0cm \mathsf{M_{a} = 30}
     \\
 \hline  
 \noalign{\smallskip}
 \mathsf{-1} \hskip 1.25cm 0.43031  \hskip 1.05cm 0.27658    \hskip 1.06cm  0.25635 \hskip 1.43cm 1.45075  \hskip 1.cm  1.03787   \hskip 1.18cm  0.94698  \\ 
 \mathsf{-0.5} \hskip 1.0cm  0.47773 \hskip 1.07cm  0.32531    \hskip 1.07cm 0.28242 \hskip 1.41cm  1.49626    \hskip 1.015cm 1.05480    \hskip 1.175cm 0.95981  \\ 
 \mathsf{0.0} \hskip 1.25cm 0.55034  \hskip 1.05cm 0.35176 \hskip 1.07cm  0.30317  \hskip 1.42cm 1.58012  \hskip 1.cm  1.06036   \hskip 1.18cm 0.94929  \\ 
 \mathsf{0.3} \hskip 1.25cm  0.62587    \hskip 1.05cm 0.37345  \hskip 1.07cm 0.31234   \hskip 1.42cm  1.67378   \hskip 1.cm   1.07176  \hskip 1.15cm 0.93804  \\ 
  \mathsf{0.5} \hskip 1.25cm  0.70256  \hskip 1.05cm  0.39530   \hskip 1.07cm 0.31871   \hskip 1.42cm   1.75753   \hskip 1.02cm 1.09095    \hskip 1.14cm 0.92967   \\ 
\hline
\noalign{\smallskip}
 \noalign{\smallskip}
\hline
 \mathsf{ \,\,\,\,\,}
 \hskip 3.5cm \rm second \hskip 0.1cm  overtone  
\hskip 4.1cm \rm third \hskip 0.1cm  overtone 
 \\
\hline
 \noalign{\smallskip}
 \noalign{\smallskip}
	 \mathsf{\alpha \,\,\,\,\,\,\,}
	 \hskip 1.1cm \mathsf{M_{a} = 3}
	 \hskip 1.0cm \mathsf{M_{a} = 10} 
	 \hskip 1.0cm \mathsf{M_{a} = 30}
	 \hskip 1.36cm \mathsf{M_{a} = 3}
 \hskip 1.0cm \mathsf{M_{a} = 10}
 \hskip 1.0cm \mathsf{M_{a} = 30}
     \\
 \hline  
 \noalign{\smallskip}
 \mathsf{-1} \hskip 1.25cm 2.50320 \hskip 1.1cm 1.81950   \hskip 1.09cm 1.68275  \hskip 1.4cm  3.55754 \hskip 0.97cm  2.59039    \hskip 1.18cm  2.40284   \\ 
 \mathsf{-0.5} \hskip 1.03cm 2.54284   \hskip 1.065cm 1.80897   \hskip 1.078cm 1.66530    \hskip 1.43cm  3.59230     \hskip 0.984cm  2.55669   \hskip 1.18cm  2.36004   \\ 
 \mathsf{0.0} \hskip 1.25cm  2.62573  \hskip 1.1cm  1.78459  \hskip 1.07cm 1.62070   \hskip 1.42cm  3.67348   \hskip 1.cm 2.50237     \hskip 1.18cm  2.28419  \\ 
 \mathsf{0.3} \hskip 1.25cm  2.72132   \hskip 1.1cm  1.77329    \hskip 1.07cm  1.58450     \hskip 1.42cm 3.76880    \hskip 1.cm  2.46542     \hskip 1.18cm  2.22368   \\   \mathsf{0.5} \hskip 1.25cm  2.79310  \hskip 1.1cm 1.77345    \hskip 1.07cm 1.55688   \hskip 1.42cm  3.82473   \hskip 0.98cm  2.43991      \hskip 1.19cm  2.17669  \\ 
\hline
\noalign{\smallskip}
\end{array}
 $$
\end{table*}
\begin{table*}
 \caption{Eigenfrequencies ($\delta_{i}$) for several  modes and  varying  magnetic field strengths for  $\gamma = 7/5$ and $\beta = 1$. }
 \label{g75b1i} 
 $$
 \begin{array}{lccccccccc}
  \noalign{\smallskip}
 \noalign{\smallskip}
\hline
 \mathsf{ \,\,\,\,\,}
 \hskip 3.5cm \rm fundamental 
\hskip 4.5cm \rm first \hskip 0.1cm  overtone 
 \\
\hline
 \noalign{\smallskip}
 \noalign{\smallskip}
	 \mathsf{\alpha \,\,\,\,\,\,\,}
	 \hskip 1.1cm \mathsf{M_{a} = 3}
	 \hskip 1.0cm \mathsf{M_{a} = 10} 
	 \hskip 1.0cm \mathsf{M_{a} = 30}
	 \hskip 1.36cm \mathsf{M_{a} = 3}
 \hskip 1.0cm \mathsf{M_{a} = 10}
 \hskip 1.0cm \mathsf{M_{a} = 30}
     \\
 \hline  
 \noalign{\smallskip}
 \mathsf{-1} \hskip 1.25cm  0.43865 \hskip 1.029cm  0.27943    \hskip 1.06cm 0.23392   \hskip 1.43cm  1.41341   \hskip 1.cm   0.84427 \hskip 1.18cm 0.69527   \\ 
 \mathsf{-0.5} \hskip 1.018cm 0.48338 \hskip 1.055cm 0.30069     \hskip 1.07cm 0.25072     \hskip 1.41cm   1.46172    \hskip 0.984cm  0.84617     \hskip 1.175cm  0.68237   \\ 
  \mathsf{0.0} \hskip 1.25cm 0.55590  \hskip 1.05cm 0.32986  \hskip 1.07cm 0.26344  \hskip 1.42cm  1.55387 \hskip 1.cm    0.85161    \hskip 1.18cm 0.65657 \\ 
 \mathsf{0.3} \hskip 1.25cm 0.62987  \hskip 1.05cm 0.35788   \hskip 1.07cm  0.27225   \hskip 1.42cm  1.64946    \hskip 1.cm  0.86141 \hskip 1.18cm   0.63397 \\ 
  \mathsf{0.5} \hskip 1.25cm 0.69568 \hskip 1.05cm 0.38442 \hskip 1.07cm 0.27884   \hskip 1.42cm  1.71604     \hskip 1.02cm  0.86861     \hskip 1.14cm   0.61193  \\ 
\noalign{\smallskip}
\hline
\noalign{\smallskip}
 \noalign{\smallskip}
\hline
 \mathsf{ \,\,\,\,\,}
 \hskip 3.5cm \rm second \hskip 0.1cm  overtone  
\hskip 4.1cm \rm third \hskip 0.1cm  overtone 
 \\
\hline
 \noalign{\smallskip}
 \noalign{\smallskip}
	 \mathsf{\alpha \,\,\,\,\,\,\,}
	 \hskip 1.1cm \mathsf{M_{a} = 3}
	 \hskip 1.0cm \mathsf{M_{a} = 10} 
	 \hskip 1.0cm \mathsf{M_{a} = 30}
	 \hskip 1.36cm \mathsf{M_{a} = 3}
 \hskip 1.0cm \mathsf{M_{a} = 10}
 \hskip 1.0cm \mathsf{M_{a} = 30}
     \\
 \hline  
 \noalign{\smallskip}
 \mathsf{-1} \hskip 1.25cm  2.43144 \hskip 1.1cm  1.44978   \hskip 1.06cm 1.19709   \hskip 1.43cm 3.45026  \hskip 1.cm 2.04889     \hskip 1.18cm  1.69333    \\ 
 \mathsf{-0.5} \hskip 1.033cm 2.47796  \hskip 1.1cm  1.42644   \hskip 1.07cm  1.14496  \hskip 1.43cm  3.49517     \hskip 0.984cm   2.00664    \hskip 1.18cm  1.61494   \\ 
 \mathsf{0.0} \hskip 1.25cm 2.57765  \hskip 1.1cm 1.39673  \hskip 1.07cm 1.06321     \hskip 1.42cm 3.60095    \hskip 1.cm 1.95155     \hskip 1.18cm 1.49445    \\ 
 \mathsf{0.3} \hskip 1.25cm  2.68125    \hskip 1.1cm  1.37676    \hskip 1.07cm  0.99033   \hskip 1.42cm  3.70963    \hskip 1.cm  1.90977    \hskip 1.18cm  1.38357   \\ 
  \mathsf{0.5} \hskip 1.25cm  2.73264   \hskip 1.1cm  1.35376    \hskip 1.07cm 0.92053    \hskip 1.42cm  3.74108   \hskip 0.98cm    1.85977    \hskip 1.19cm  1.27176   \\ 
\noalign{\smallskip}
\hline
\noalign{\smallskip}
\end{array}
 $$
\end{table*}
\begin{table*}
 \caption{Eigenfrequencies ($\delta_{i}$) for the various modes of  parameters $\gamma = 7/5$ and $\beta = 2$ for various magnetic fields. }
 \label{g75b2i} 
 $$
 \begin{array}{lccccccccc}
  \noalign{\smallskip}
 \noalign{\smallskip}
\hline
 \mathsf{ \,\,\,\,\,}
 \hskip 3.5cm \rm fundamental 
\hskip 4.5cm \rm first \hskip 0.1cm  overtone 
 \\
\hline
 \noalign{\smallskip}
 \noalign{\smallskip}
	 \mathsf{\alpha \,\,\,\,\,\,\,}
	 \hskip 1.1cm \mathsf{M_{a} = 3}
	 \hskip 1.0cm \mathsf{M_{a} = 10} 
	 \hskip 1.0cm \mathsf{M_{a} = 30}
	 \hskip 1.36cm \mathsf{M_{a} = 3}
 \hskip 1.0cm \mathsf{M_{a} = 10}
 \hskip 1.0cm \mathsf{M_{a} = 30}
     \\
 \hline  
 \noalign{\smallskip}
 \mathsf{-1} \hskip 1.25cm 0.42613  \hskip 1.06cm  0.24811    \hskip 1.06cm 0.19424   \hskip 1.43cm 1.41061    \hskip 0.98cm  0.84868 \hskip 1.18cm 0.71088   \\

 \mathsf{-0.5} \hskip 1.018cm 0.46608 \hskip 1.055cm 0.26937     \hskip 1.07cm 0.21581      \hskip 1.41cm 1.45638   \hskip 0.984cm   0.85532     \hskip 1.175cm  0.71608  \\

 \mathsf{0.0} \hskip 1.25cm 0.53258  \hskip 1.05cm 0.29198   \hskip 1.07cm 0.22668  \hskip 1.42cm 1.54505  \hskip 1.cm  0.86521      \hskip 1.18cm  0.70578 \\

 \mathsf{0.3} \hskip 1.25cm   0.60136 \hskip 1.05cm  0.31426  \hskip 1.07cm 0.23328    \hskip 1.42cm 1.63895     \hskip 1.cm 0.88205  \hskip 1.18cm  0.69826  \\ 

  \mathsf{0.5} \hskip 1.25cm 0.66468  \hskip 1.05cm  0.33806 \hskip 1.07cm 0.23891   \hskip 1.42cm    1.70814   \hskip 1.02cm 0.90367       \hskip 1.14cm 0.69442   \\ 

\hline
\noalign{\smallskip}
 \noalign{\smallskip}
\hline
 \mathsf{ \,\,\,\,\,}
 \hskip 3.5cm \rm second \hskip 0.1cm  overtone  
\hskip 4.1cm \rm third \hskip 0.1cm  overtone 
 \\
\hline
 \noalign{\smallskip}
 \noalign{\smallskip}
	 \mathsf{\alpha \,\,\,\,\,\,\,}
	 \hskip 1.1cm \mathsf{M_{a} = 3}
	 \hskip 1.0cm \mathsf{M_{a} = 10} 
	 \hskip 1.0cm \mathsf{M_{a} = 30}
	 \hskip 1.36cm \mathsf{M_{a} = 3}
 \hskip 1.0cm \mathsf{M_{a} = 10}
 \hskip 1.0cm \mathsf{M_{a} = 30}
     \\
 \hline  
 \noalign{\smallskip}
 \mathsf{-1} \hskip 1.25cm   2.42930 \hskip 1.1cm 1.47335    \hskip 1.06cm  1.25896  \hskip 1.43cm 3.44982  \hskip 1.cm   2.09150      \hskip 1.18cm 1.79406   \\

 \mathsf{-0.5} \hskip 1.033cm 2.47396  \hskip 1.08cm   1.46111   \hskip 1.07cm 1.23842   \hskip 1.43cm  3.49352      \hskip 0.984cm 2.06317    \hskip 1.18cm 1.75504   \\

 \mathsf{0.0} \hskip 1.25cm 2.57015   \hskip 1.1cm 1.44656   \hskip 1.07cm  1.20060    \hskip 1.42cm   3.59567     \hskip 1.cm   2.02658     \hskip 1.18cm 1.69333    \\

 \mathsf{0.3} \hskip 1.25cm  2.67207     \hskip 1.1cm   1.44242   \hskip 1.07cm   1.17200    \hskip 1.42cm 3.70268     \hskip 1.cm    2.00303   \hskip 1.18cm 1.64574    \\ 

  \mathsf{0.5} \hskip 1.25cm   2.72750   \hskip 1.1cm  1.44027      \hskip 1.07cm 1.15139     \hskip 1.42cm   3.73955   \hskip 0.98cm  1.97825     \hskip 1.19cm    1.60978 \\

\noalign{\smallskip}
\hline
\noalign{\smallskip}
\end{array}
 $$
\end{table*}

\label{lastpage}
\end{document}